\documentclass[letterpaper,twocolumn,10pt]{article}
\usepackage{usenix}

\usepackage{graphicx}
\usepackage{subfigure}
\usepackage{amsmath}
\usepackage{amsthm}
\usepackage{amssymb}
\usepackage{mathrsfs}
\usepackage{float}
\usepackage[vlined,ruled,linesnumbered]{algorithm2e}
\usepackage{xcolor}
\usepackage{multicol}
\usepackage{multirow}

\begin{document}
%
\title{Model Inversion Attack against Transfer Learning: Inverting a Model without Accessing It}


\author{
{\rm Dayong Ye}\\
UTS, Australia\\
Dayong.Ye@uts.edu.au
\and
{\rm Huiqiang Chen}\\
UTS, Australia\\
Huiqiang.Chen@student.uts.edu.au
\and
{\rm Shuai Zhou}\\
UTS, Australia\\
Shuai.Zhou@student.uts.edu.au
\and
{\rm Tianqing Zhu}\\
UTS, Australia\\
Tianqing.Zhu@uts.edu.au
\and
{\rm Wanlei Zhou}\\
City University of Macau, China\\
wlzhou@cityu.mo
\and
{\rm Shouling Ji}\\
Zhejiang University, China\\
sji@zju.edu.cn
\and
}

\maketitle

\begin{abstract}
Transfer learning is an important approach that produces pre-trained teacher models which can be used to quickly build specialized student models. 
However, recent research on transfer learning has found that it is vulnerable to various attacks, e.g., misclassification and backdoor attacks. 
However, it is still not clear whether transfer learning is vulnerable to model inversion attacks. 
Launching a model inversion attack against transfer learning scheme is challenging. 
Not only does the student model hide its structural parameters, but it is also inaccessible to the adversary. 
Hence, when targeting a student model, both the white-box and black-box versions of existing model inversion attacks fail. 
White-box attacks fail as they need the target model's parameters. 
Black-box attacks fail as they depend on making repeated queries of the target model. 
However, they may not mean that transfer learning models are impervious to model inversion attacks. 
Hence, with this paper, we initiate research into model inversion attacks against transfer learning schemes with two novel attack methods. 
Both are black-box attacks, suiting different situations, that do not rely on queries to the target student model. 
In the first method, the adversary has the data samples that share the same distribution as the training set of the teacher model. 
In the second method, the adversary does not have any such samples. 
Experiments show that highly recognizable data records can be recovered with both of these methods. 
This means that even if a model is an inaccessible black-box, it can still be inverted. 
\end{abstract}

\section{Introduction}\label{sec:introduction}
The success of machine learning has been the driving force behind a wide variety of applications, from computer vision to speech recognition \cite{Ye15,Salem19}. 
Thus, due to their power and place in model software systems, 
machine learning models, like deep neural networks, have become important pieces of intellectual property for their owners 
due to their powerful capacity in modern software systems \cite{Amazon}. 
However, building a powerful machine learning model requires both a massive amount of data and enormous computational resources. 
Further, this hunger for data coupled with the exorbitant computational overhead limits the number of models that can be independently trained. 
One solution to this problem has been transfer learning \cite{Zhuang21}, 
where a small number of well-trained teacher models is shared with the general community. 
Individual users then can customize these teacher models, building student models to suit their particular tasks 
with a much smaller amount of data and far less computational resources. 

Although transfer learning is able to address the data scarcity problem, 
its open nature attracts a diversity of adversaries 
who exploit teacher models to compromise individual student models. 
Research shows that adversarial activities against transfer learning, 
include both misclassification attacks \cite{Wang18} and membership inference attacks \cite{Zou20}. 
However, among these adversarial activities, a common type of attack has not been investigated - that being model inversion attacks. 
The aim of a model inversion attack is to reconstruct a model's input data 
based on its outputs (black-box) or model parameters (white-box) \cite{Fred14,Fred15}. 
Existing model inversion attacks have mainly been mounted against standalone models \cite{Yang19,Zhang20CVPR} not transfer learning models. 
That, however, does not necessarily mean that model inversion attacks against transfer learning models are not possible. 
Hence, to explore this potential, we examined whether transfer learning is vulnerable to model inversion attacks. 

The research of model inversion against transfer learning is challenging, 
because 1) student models are private and cannot be accessed by adversaries; 
and 2) this privateness makes it difficult for an adversary to collect a great many samples with the same distribution as the student model's training set. 
Existing white-box attacks \cite{Geiping20,Yin21} require the target model's parameters, e.g., the architecture or the gradients. 
Such information about the target student model, however, is not available in transfer learning. 
Although existing black-box attacks require only the target model's outputs \cite{Yang19,Zhao21}, 
they can typically only acquire this information through a large number of queries to the target model. 
As such, they assume that collecting a massive amount of data sharing the same distribution as the target model's training set is possible \footnote{Although Yang et al. \cite{Yang19} also used general distribution auxiliary data, their major results were derived from the same distribution data.}. 
On the surface, it would therefore seem that transfer learning models are robust to model inversion attacks. 
However, we contend that such data and queries may not be required and propose two novel model inversion attack methods 
that do not require access to the target model. 

In the first method, an adversary is assumed to have a set of data that has the same distribution as the training data of the teacher model. 
She first uses these data to train an inversion model against the teacher model, 
and then trains a conversion model to assist the inversion model to suit the target student model using a limited amount of data. 
These data share the same distribution as the training data of the student model. 
In the second method, the adversary is assumed not to have any teacher data. 
She first collects an amount of unlabeled data that have a similar distribution to the training data of the teacher model. 
Then, she augments her dataset using these unlabeled data. 
The augmented dataset is used to train a shadow model to mimic the target student model. 
It is also used to train an attack model to invert the shadow model. 
As the shadow model is similar to the target student model, 
the attack model can also be used to invert the student model. 
In summary, we make the following three key contributions.

    1) To the best of our knowledge, we are the first to investigate the vulnerability of transfer learning under model inversion attacks. 
    
    2) To overcome the issue of the target student model being inaccessible and the problem of the amount of available data being limited, 
    we propose two novel attack methods, each of which suits a distinct situation. 
    
    3) We extensively evaluate the effectiveness of the two methods in diverse settings, 
    including both facial images and hand-drawn digits. 

\section{Related Work}\label{sec:related work}
This section reviews the related research of transfer learning attacks and model inversion attacks, 
highlighting how our research differs from existing research. 

\subsection{Transfer Learning Attacks}
Existing transfer learning attacks can be classified into four categories: 
misclassification attacks, membership inference attacks, backdoor attacks and model extraction attacks.

\subsubsection{Misclassification Attacks}
This category of attacks aims to have a student model misclassify targeted inputs to specified classes. 

Wang et al. \cite{Wang18} proposed a misclassification attack method by adding perturbations to source images 
to mimic the internal representation of target images at layer $K$ of the teacher model. 
The key insight is that if a perturbed source image's internal representation at layer $K$ matches that of the target image, 
the perturbed source image will be misclassified into the same class as the target image.  

Ji et al. \cite{Ji18} also designed a misclassification attack method that uses the teacher model as a genuine feature extractor. 
Unlike \cite{Wang18} which perturbs the source images, Ji et al. created an adversarial model 
by slightly modifying a subset of the teacher model's parameters. 

Rezaei and Liu \cite{Rezaei20} developed a brute force attack that can craft instances of input to trigger each target class. 
Their idea is to activate the $i^{th}$ neuron at the output of teacher model's $(N-1)^{th}$ layer 
while keeping the other neurons at the same layer zero. 
Then, 
if there exists a neuron at the $N^{th}$ layer that associates a large weight to the activated neuron, that neuron will become large. 
Hence, an adversary image can be created by iteratively triggering each neuron at the $N^{th}$ layer.



\subsubsection{Membership Inference Attacks}
This category of attacks aims to infer whether a given data record is in the training set of a student or teacher model 
with the assistance of the other model. 


Hidano et al. \cite{Hidano21} developed a transfer shadow training technique 
that uses the parameters of a student model to construct shadow models. 
These shadow models are used to create an attack training set for a learning-based adversary, 
or to compute the modified prediction entropy for each sample from a shadow set for an entropy-based adversary. 

\subsubsection{Backdoor Attacks}
This category of attacks first trains a trigger, usually a small pattern in image classification problems. 
Then, once the trigger is activated, the victim model will classify any given sample to a desired class. 

Yao et al. \cite{Yao19} 
embedded a latent trigger on a non-existent output label 
and had the victim activate the backdoor themselves when they performed transfer learning. 
Their latent attacks associate a trigger with an intermediate layer of the teacher model. 
Then, for any victim who uses this infected teacher model for transfer learning, 
the student model will also be infected by attaching the trigger.

Wang et al. \cite{Wang20TSC} manipulated the teacher model to generate customized student models that give wrong predictions. 
They first selected a set of neurons from the teacher model based on their importance. 
Then, they generated a trigger by minimizing the distortion between the reconstructed input and its backdoored version. 
Finally, they crafted the backdoored model by retraining the genuine model using the malicious input incorporated with the trigger.

\subsubsection{Model Extraction Attacks}
This category of attacks aims at discovering the parameters of a target model, typically the structure and weights. 

Breier et al. \cite{Breier21} presented an attack method to precisely recover the parameters of the last hidden layer of the student model. 
They change the intermediate values of the student model 
by intentionally injecting faults into the model. 
Then, they observe the corresponding change of the model's output, 
which reveals information about the model's parameters. 

\subsection{Model Inversion Attacks}
Existing model inversion attacks can be classified into two categories: black-boxed and white-boxed. 
In black-box attacks, the adversary can only observe the output of the target model, 
while in white-box attacks, the adversary is aware of all the information of the target model. 

\subsubsection{Black-box Model Inversion Attacks}
Fredrikson et al. \cite{Fred14} initiated the research of model inversion attacks 
on personalized Warfarin dosing in a black-box manner. 
Their method works by estimating the probability of a potential target attribute given the available information and the model. 


Unlike Fredrikson et al.'s optimization-based methods, 
Yang et al.'s method \cite{Yang19} is training-based. It inverts a model by learning a second model that acts as the inverse of the original one. 
The second model takes the predicted confidence vectors of the original model as input, and outputs reconstructed data. 
Zhao et al. \cite{Zhao21} improved Yang et al.'s work \cite{Yang19} by incorporating image explanations into the target model's output vectors. 
The explanations and vectors are then used together to train an inversion model. 
Although Zhao et al.'s method outperforms Yang et al.'s method, Zhao et al's method requires that either the target model can output image explanations 
or the attacker can train a surrogate model with the same functions as the target model. 
However, neither of these requirements is feasible in our problem. 


In addition to the prevalent setting, model inversion attacks have also been studied in other settings. 
Salem et al. \cite{Salem20} studied data reconstruction attacks in an online learning setting, 
while Carlini et al. \cite{Carlini21} studied data extraction from language models.


\subsubsection{White-box Model Inversion Attacks}

Zhang et al. \cite{Zhang20} presented a generative model inversion attack method 
that can invert DNNs and synthesize private training data with high fidelity. 
Their method involves two stages: public knowledge distillation and secret revelation. 
In the first stage, a generator and multiple discriminators 
are trained on public datasets to encourage the generator to create visually realistic images. 
In the second stage, the sensitive regions in the images created in the first stage 
are recovered by solving an optimization problem.

Geiping et al. \cite{Geiping20} investigated model inversion attacks in federated learning settings using users' gradients 
where the server separately stores and process individual user gradients. 
By using the norm magnitude and direction contained in gradients, 
the server is able to reconstruct high-quality data records. 
Yin et al. \cite{Yin21} improved Geiping et al.'s method \cite{Geiping20} by proposing a new method named GradInversion. 
GradInversion can recover a batch of images using average gradients. 

\subsection{Summary of Related Work}
Current related research on transfer learning attacks overlooks model inversion attacks, 
while current model inversion attacks are only launched against standalone models. 
We aim to fill this gap by exploring model inversion attacks against transfer learning. 

\section{Preliminary}\label{sec:preliminary}
\subsection{Transfer Learning}
The key idea of transfer learning is to transfer the knowledge from a pre-trained teacher model to a new student model, 
where the two models share a significant similarity \cite{Zhuang21}. 
The knowledge is typically contained in the parameters of the teacher model, e.g., the structure and weights of a deep neural network. 
Hence, the knowledge can be transferred by freezing most layers of the teacher model 
while only fine-tuning the last few layers to produce a student model. 

\begin{figure}[ht]
\centering
	\includegraphics[scale=0.6]{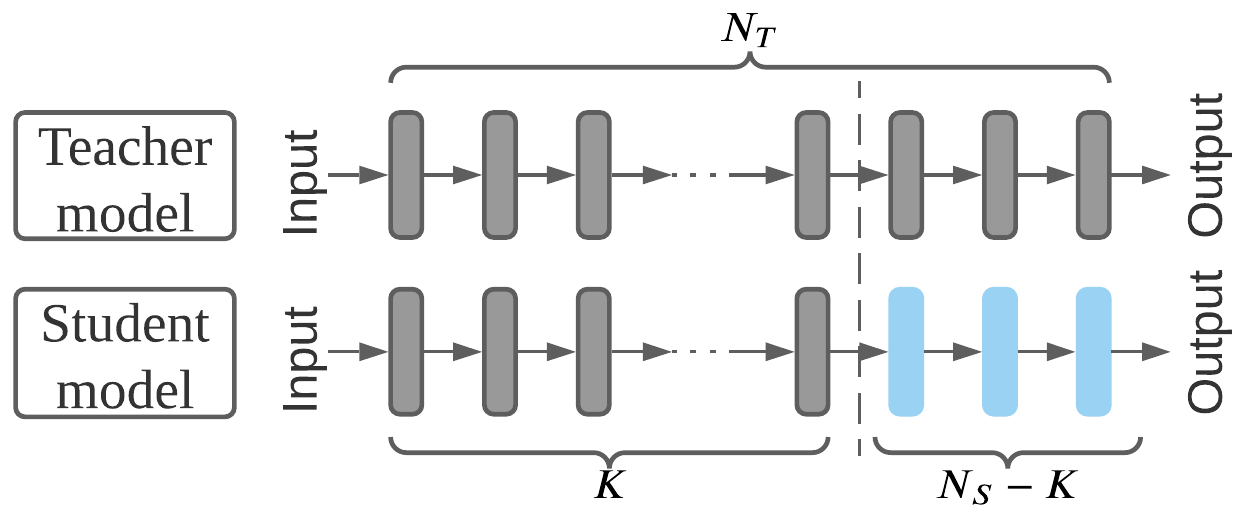}
	\caption{Transfer learning. A student model $S$ copies and freezes the first $K$ layers from a teacher model $T$, 
    updating only the remaining $N_S-K$ newly added layers.} 
	\label{fig:TransferLearning}
\end{figure}

Let $T$ be the teacher model with $N_T$ layers and let $S$ be the student model with $N_S$ layers. 
In Figure \ref{fig:TransferLearning}, the student model $S$ is initialized by copying the first $K$ layers of the teacher model $T$. 
Then, the weights of the first $K$ layers are fixed while the weights of the remaining $N_S-K$ layers are updated using the student's dataset. 
The first $K$ layers are frozen during training because those layers can extract the general features of the input samples. 
Given that the teacher and student models are highly similar, these extracted features can be directly used by the student model, 
which reduces the training time and the amount of training data required. 
Based on the number of layers being frozen, $K$, 
transfer learning can be divided into the following three categories \cite{Wang18}.

    1) Deep-layer feature extractor ($K=N_T-1$): The first $N_T-1$ layers are frozen and only the last classification layer is updated. 
    This extractor is most often used when the student task is very similar to the teacher task, e.g., both tasks are about facial recognition.
    
    2) Mid-layer feature extractor ($0<K<N_T-1$): The first $K$ layers are frozen. 
    This extractor allows more layers to be tuned, which enables the knowledge to be transferred  
    even if the student task is slightly different from the teacher task. 
    For example, the student task is about facial recognition while the teacher task is about traffic sign recognition.
    
    3) Full model fine-tuning ($K=0$): All the layers are unfrozen and have to be tuned during the training. 
    This extractor is applied to situations where the student task is significantly different from the teacher task. 
    As discussed in \cite{Weiss16}, even if the difference is huge, 
    using this extractor enables the student model to converge faster and achieve better performance than training from scratch. 
    Here, for example, the student task may be about facial recognition while the teacher task is about flower recognition.

\subsection{Model Inversion Attacks}
Model inversion attacks aim to reconstruct a target model's input data from its output by inverting the target model \cite{Fred15}.
Typically, the adversary trains a new attack model on an auxiliary dataset, 
which acts as the inverse of the target model \cite{Yang19}. 
The attack model takes the output of the target model as input 
and tries to reconstruct the original input data.

Formally, let $F$ be the target model consisting of $m$ classes, and let $G$ be the attack model. 
Given a data record $(x,y)$, where $x$ is the data and $y$ is the label of $x$, 
the adversary inputs $x$ into $F$ and receives $F(x)$. 
$F(x)$ is then fed into $G$ which returns $G(F(x))$ that is expected to be very similar to $x$. 
In other words, the aim is to minimize the following objective: 
\begin{equation}
    C(G)=\mathbb{E}_{x\sim p_{x}}[L(G(F(x)),x)],
\end{equation}
where $p_{x}$ is the underlying probability distribution of $x$, and $L$ is a loss function such as an $L_2$ loss.

\subsection{Threat Model}
Given a teacher model $T$ and a student model $S$, the adversary can access the teacher model $T$ in a white-boxed manner, 
namely the adversary knows the structure and parameters of model $T$, and can observe the output of each layer in model $T$. 
The adversary also has a set of samples with the same distribution as the teacher model $T$'s training data. 
This set of samples is called the ``teacher data'' and denoted as $D_T$.
This assumption will be relaxed in our second method. 

The adversary cannot access the student model $S$ but can observe the output confidence scores of model $S$. 
The adversary has a limited number of samples (less than $10$ for each class) 
which share the same distribution as the training data of model $S$. 
These samples are called the ``student data'' and denoted as $D_S$. 
By comparison, in most existing research, e.g., \cite{Yang19,Zhao21}, it is assumed that the adversary had a large number of samples 
with the same distribution as the target model's training data. 
In extreme situations where the adversary does not have any student data with which 
to successfully launch an inversion attack, the adversary must have external knowledge regarding 
the semantic relationships between the teacher model $T$'s classes and the student model $S$'s classes. 
However, given that the adversary cannot access the student model $S$, quantifying these semantic relationships may be intractable. 
Zero-shot learning \cite{Wang19,Chen21} may offer some ideas to address this issue, 
but it aims to predict unseen classes rather than inverting a given model. 
Thus, we leave research on this extreme situation to future work. 

In addition, the adversary knows which teacher model $T$ was used to build the student model $S$ and 
which transfer learning method was used to train model $S$. 
This information is easy to obtain, because many service providers, e.g., Facebook PyTorch \cite{PyTorch}, 
offer such information in their official tutorials. 
The aim of the adversary is to invert the student model $S$ 
so that for each input sample to $S$, the adversary can reconstruct this sample based on the output of $S$. 


A motivating example comes from a small hospital which has only a small dataset of X-ray images and 
wants to achieve high performance of image classification for accurate diagnosis. 
The small hospital uses transfer learning by exploiting a teacher model from a large hospital or public image classification models. 
For an adversary who can observe the output of the small hospital's model but cannot access it, 
she can still use our method to invert that model. 
Then, for any individual patient, once the adversary has the model's output regarding that patient, she can reconstruct that patient's X-ray image, 
which breaches that patient's privacy. 
Note that the output, itself, may be meaningless to the adversary, as it could simply be a real-number vector. 
Hence, to view the patient's health status, the adversary must recover the X-ray image from the output vector.


\section{The Design of Attack Methods}\label{sec:method}
As mentioned, there are two challenges in designing model inversion attacks on transfer learning models, 
i.e., inaccessibility to the target student model and having only a limited amount of student data. 
In the first method, the adversary first trains an attack model to invert the teacher model.
Then, she uses the limited student data to transfer the attack model to invert the student model. 
The rationale is that as the adversary has a number of teacher data, 
she can train a teacher inversion model well. 
Moreover, as the adversary has some student data, she can transfer this teacher inversion model to a student inversion model. 
In this process, no query to the student model is required. 
Also, to transfer a well-trained teacher inversion model to a specific student inversion model, a limited amount of student data is enough. 

In the second method, as the adversary does not have any teacher data, she cannot invert the teacher model. 
Thus, she trains a shadow model to mimic the student model. 
Then, she trains an attack model to invert the shadow model. 
After that, the attack model is used to reconstruct the inputs of the student model. 
The rationale is that the adversary has only a limited amount of student data, 
but she can collect a large amount of unlabeled data which have similar distribution to the student data. 
Thus, she can augment her dataset which is then used to train a shadow model, having similar features to the student model. 
Then, she uses the augmented dataset to train an inversion model against this shadow model. 
As the shadow model and the student model are similar, the inversion model can also be used to invert the student model. 
Again, in this process, no query to the student model is needed. 
Also, due to the use of data augmentation technique, a limited amount of student data is enough. 

One may think that, as the adversary knows all the information of the teacher model and shares a part of parameters with the student model, 
then the adversary can easily use existing white-box attack methods to invert the student model. This, however, is not the case. 
The main reason is that white-box attacks, e.g., \cite{Geiping20,Yin21}, typically need the gradients of the target model. 
To compute the gradients, the adversary has to know the loss function used by the target model. 
This information, however, is unknown to the adversary in our setting. 

\subsection{Design of the First Attack Method}
\subsubsection{Overview of the First Method}
\begin{figure}[ht]
\centering
	\includegraphics[scale=0.38]{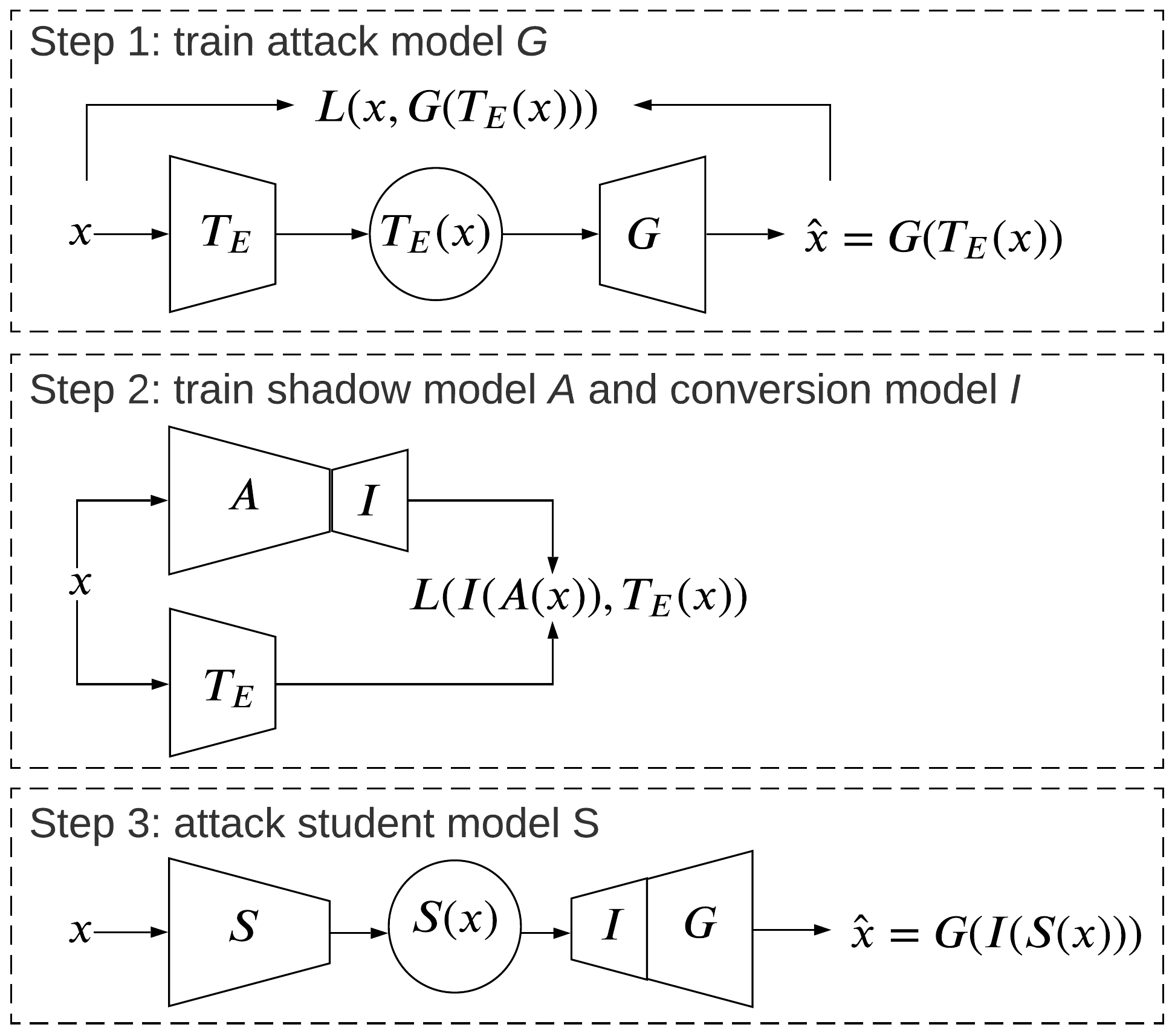}
	\caption{Overview of the first method. In Step 1, The adversary trains an attack model $G$ exploiting the modified teacher model $T_E$. In Step 2, the adversary crafts a shadow model $A$, based on the teacher model $T$, and creates a conversion model $I$. Then, models $A$ and $I$ are connected for training. In Step 3, the adversary connects model $I$ with model $G$ to invert the target student model $S$.} 
	\label{fig:overview1}
\end{figure}

Figure \ref{fig:overview1} shows the overview of the first attack method which consists of the following three steps. 

\noindent\textbf{Step 1.} The adversary trains an attack model to invert the teacher model 
using the samples which have the same distribution as the teacher model's training set, i.e., teacher data. 

\noindent\textbf{Step 2.} The adversary builds a shadow model and a conversion model.  
The two models are connected for training using both the teacher data and a limited amount of student data.  

\noindent\textbf{Step 3.} The adversary connects the conversion model with the target student model to launch the model inversion attack. 

\subsubsection{Step 1: Attack Model Training}
In the first method, we assume that the adversary has a set of samples, $D_T$, that shares the same distribution as the training data of the teacher model. 
Hence, the adversary can use $D_T$ to train an attack model $G$ against the teacher model $T$. 
The process of training model $G$ is as follows. 
First, the adversary removes the fully-connected layers from the teacher model $T$ while keeping only the CNN layers. 
The modified teacher model is denoted as $T_E$, which is used as a feature extractor. 
Then, the adversary uses the teacher data $D_T$ to train the attack model $G$. 
Specifically, the modified teacher model $T_E$ takes a sample $x$ from $D_T$ as input 
and produces a vector $T_E(x)$ that is fed to model $G$ which outputs the reconstructed data $\hat{x}=G(T_E(x))$. 
The loss function is mean squared error. 

Note that a regular training process is not used to train the attack model $G$ in that 
the process does not take the teacher model's last layer's output as input \cite{Yang19}. 
This is because the model's last layer only holds classification information 
whereas the previous layers hold much more information. 
Hence, directly using the output of the CNN layers to train the attack model $G$ yields better reconstruction quality.

\subsubsection{Step 2: Shadow Model and Conversion Model Training}
After the attack model $G$ is trained, the adversary needs to make $G$ suit the student model. 
However, this is a challenging task. 
The dimension of model $G$'s input is the same as that of the modified teacher model $T_E$'s output. 
This dimension, however, is not usually identical to the student model $S$'s output. 
Therefore, the adversary has to train a conversion model $I$ to map model $S$'s output to model $G$'s input. 

To train model $I$, the adversary first builds a shadow model $A$. 
Model $A$ is transferred from the teacher model $T$ by replacing the last layer of model $T$ with a new output layer 
whose dimension is identical to the student model $S$'s output dimension. 
Then, the adversary connects models $A$ and $I$ together, 
and uses both the teacher data $D_T$ and student data $D_S$ to train them. 
During the training, the adversary inputs each sample $x$ from $D_T$ and $D_S$ into model $A$, 
and receives the output of model $I$ denoted as $I(A(x))$. 
The adversary also inputs sample $x$ into the modified teacher model $T_E$ and receives the output $T_E(x)$, 
which is used as ground truth to update models $A$ and $I$. 
The loss function is the mean squared error $L(I(A(x)),T_E(x))$. 
The reason for including the teacher data to update the two models is explained in the next section. 


Note that, at this point, the shadow model $A$ and the conversion model $I$ must be connected so that they can be trained together, rather than training them separately. 
This is because model $A$ is only used to encode a sample to a vector  
with a dimensionality that is identical to that of the student model $S$'s output. 
If models $A$ and $I$ were to be trained separately, then model $A$ would be trained as a classifier 
and we could only use the student data $D_S$ to train it. 
However, in our assumptions, the adversary only has a limited amount of student data 
that are insufficient to train model $A$ well. 

\subsubsection{Step 3: Invert the Target Student Model}
Once the training of models $I$ and $A$ is complete, the adversary connects the conversion model $I$ with the attack model $G$. 
Then, for any output vectors from the student model $S$, i.e., $S(x)$, 
the adversary inputs $S(x)$ into models $I$ and $G$, 
and receives a reconstructed sample $\hat{x}=G(I(S(x)))$.

\subsection{Design of the Second Attack Method}
The first attack method requires an amount of teacher data. 
However, in the real-world, teacher data may not always be available, 
which gives rise to our second attack method - one that does not depend on any teacher data. 

\subsubsection{Overview of the Second Method}
\begin{figure}[ht]
\centering
	\includegraphics[scale=0.4]{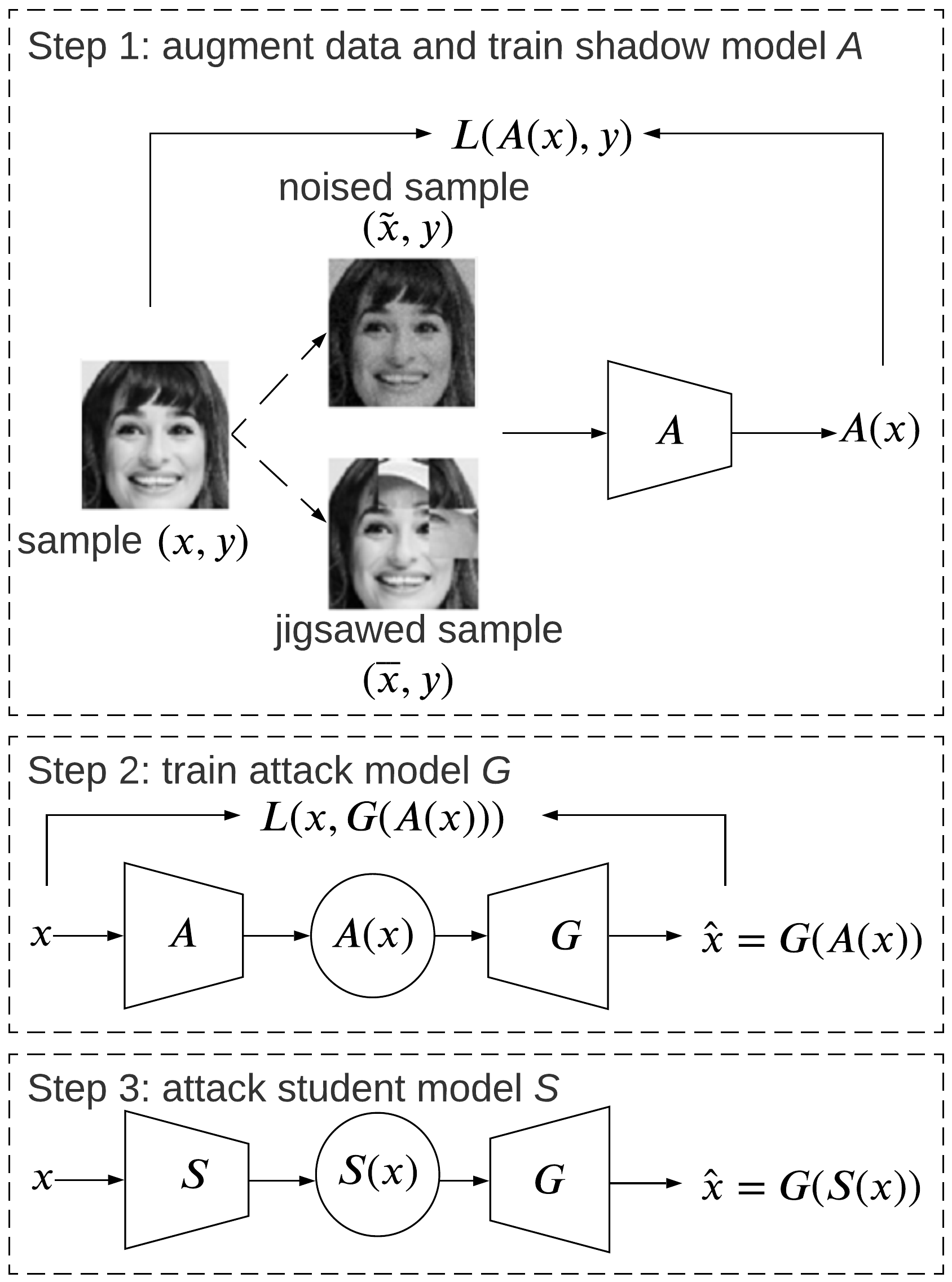}
	\caption{Overview of the second method. In Step 1, the adversary augments data using both the noise-based and jigsaw-based approaches, and then uses the augmented data to train a shadow model $A$. In Step 2, the adversary trains an attack model $G$ to invert model $A$. In Step 3, the adversary uses model $G$ to attack the target student model $S$.} 
	\label{fig:overview2}
\end{figure}

Figure \ref{fig:overview2} shows an overview of the second attack method, which consists of the following three steps.  

\noindent\textbf{Step 1.} The adversary augments the data and trains a shadow model.

\noindent\textbf{Step 2.} The adversary trains an attack model to invert the shadow model. 

\noindent\textbf{Step 3.} The adversary uses this attack model to reconstruct input samples of the student model. 

\subsubsection{Step 1: Data Augmentation and Shadow Model Training}
Data augmentation techniques have been widely used in deep learning to address the problem of limited data \cite{Shorten19}. 
These techniques have also been combined with one-shot learning which aims to learn classifiers with only a few labeled examples of each class \cite{Wang20}. 
Here, we also use one-shot learning with data augmentation to train a shadow model. 

We adopt two data augmentation approaches for one-shot learning: 
the noise-based approach \cite{Fong17,Kurakin17,Ji18} and the jigsaw-based approach \cite{Chen19,Yun19}. 
The rationale for the noise-based approach is that adding proper noise to a given sample can generate its ``neighboring'' samples 
with the same main features as the given sample. 
Further, the jigsaw-based approach is able to generate the samples that intentionally destroy the spatial coherence of the given sample \cite{Sinha21}. 
Thus, the jigsawed samples typically sit at a large distance away from the given sample in the data space. 
These generated samples are expected to contain various features of the class to which the given sample belongs. 
Hence, using these samples could train a shadow model to mimic the student model.

\vspace{2mm}
\noindent\textbf{The noise-based data augmentation approach} \hspace{3mm} Formally, let $x^*_c$ be a given sample for class $c$, 
where class $c$ is one of the student model $S$'s classes and $x^*_c$ is possessed by the adversary. 
To add proper noise to $x^*_c$, the adversary carefully crafts noise to make the noisy samples semantically similar to $x^*_c$,
namely these noisy samples can be classified as the same class as $x^*_c$. 
To achieve this goal, the adversary first feeds $x^*_c$ to the teacher model $T$ 
and receives the output of the $K$th layer of model $T$ denoted as $T_K(x^*_c)$. 

The adversary then introduces a mask $m_c$ for $x^*_c$ \cite{Fong17}. 
Each dimension $i$ of $x^*_c$, i.e., $x^*_c[i]$, is associated with a scalar $m_c[i]\in[0,1]$. 
To generate a mask $m_c$, the adversary defines a perturbation operator $\psi$: 
\begin{equation}\label{eq:perturb}
    \psi(x^*_c;m_c)[i]=m_c[i]\cdot x^*_c[i]+(1-m_c[i])\cdot\eta,     
\end{equation}
where $\eta$ is a random noise sampled from Gaussian distribution $\mathcal{N}(0,\sigma^2)$. 
Thus, when $m_c[i]=1$, no noise is added to $x^*_c[i]$; when $m_c[i]=0$, some random noise replaces $x^*_c[i]$. 
The adversary aims to find $m_c$ such that $x^*_c$'s main features can be preserved while those remaining features are perturbed. 
To this end, the adversary defines a learning problem: 
\begin{equation}\label{eq:mask}
    m^*_c=\mathop{argmin}\limits_{m_c}D(T_K(\psi(x^*_c;m_c)),T_K(x^*_c))-\lambda\cdot||m_c||_1, 
\end{equation}
where the first term $D(\cdot,\cdot)$ measures the Euclidean distance between two vectors, 
and the second term $\lambda\cdot||m_c||_1$ encourages most of the mask to be close to $1$, i.e., perturbing a small part of $x^*_c$. 
The parameter $\lambda$ balances the two terms. 
In particular, the learning problem tries to minimize the distance between a generated sample $\psi(x^*_c;m_c)$ and the given sample $x^*_c$. 
Note that the adversary is not minimizing the direct distance between the two samples. 
Instead, she feeds the two samples into the teacher model $T$ 
and receives the outputs of the $K$th of model $T$, i.e., $T_K(\psi(x^*_c;m_c))$ and $T_K(x^*_c)$. 
The adversary then minimizes the distance between $T_K(\psi(x^*_c;m_c))$ and $T_K(x^*_c)$. 
We reason that if two samples can be classified as the same class by the student model $S$, they must have very similar features. 
As the student model $S$ is trained by transferring parameters from the teacher model $T$, 
model $T$ can be used as a feature extractor whose output of the $K$th layer contains the features of the input sample. 
Once the mask $m_c$ is created, the adversary uses the perturbation operator $\psi$ to generate a set of neighboring samples $\psi(x^*_c;m_c)$ of $x^*_c$. 

\vspace{2mm}
\noindent\textbf{The jigsaw-based data augmentation approach} \hspace{3mm} 
In addition to neighboring samples, the adversary also needs extra samples 
that are far away from, but still in the same class as, the given sample $x^*_c$. 
For this, the adversary uses the jigsaw-based approach. 
Concretely, the adversary collects a set of unlabeled auxiliary samples, $D_A$, which share their main features with $x^*_c$. 
For example, if $x^*_c$ is a woman's facial image, the auxiliary samples may also be images of women's faces but not the same person. 
The adversary, then, resizes these auxiliary samples to the same size as $x^*_c$. 
Next, she introduces a binary mask $M_c\in\{0,1\}^{W\times H}$ indicating where to drop out and fill in from two images. 
Here, $W$ and $H$ denote the width and height of $x^*_c$, respectively. 
The mask $M_c$ is used to mix the two samples. 
Formally, given a sample $x$ from the auxiliary set $D_A$, the mixed sample $\tilde{x}^*_c$ is computed as 
\begin{equation}
    \tilde{x}^*_c=M_c\odot x^*_c+(\mathbf{1}-M_c)\odot x, 
\end{equation}
where $\mathbf{1}=\{1\}^{W\times H}$ is a binary mask filled with ones and $\odot$ is element-wise multiplication. 
The adversary aims to find an $M_c$ such that the mixed sample $\tilde{x}^*_c$ has similar features to $x^*_c$. 
To this end, the adversary defines a learning problem: 
\begin{equation}\label{eq:mask2}
    M^*_c=\mathop{argmin}\limits_{M_c}D(T_K(\tilde{x}^*_c),T_K(x^*_c)).
\end{equation}

Here, the mask generated using Eq. \ref{eq:mask2} is similar to Eq. \ref{eq:mask}. 
The main difference is that Eq. \ref{eq:mask2} does not include the second term as in Eq. \ref{eq:mask}, $\lambda\cdot||m_c||_1$. 
This means that Eq. \ref{eq:mask2} does not encourage using a small part of an auxiliary sample $x$ to mix with $x^*_c$, 
but, rather, focuses only on the features of the mixed sample $\tilde{x}^*_c$ and the given sample $x^*_c$. 
Hence, to a large extent, Eq. \ref{eq:mask2} avoids generating neighboring samples of $x^*_c$. 


Finally, the noisy samples and jigsawed samples are labelled as $c$ and stored in a dataset, denoted as $D^c_{aug}$. 
The adversary then uses the same approaches to generate noisy and jigsawed samples for the other classes of the student model $S$. 
When a dataset $D^c_{aug}$ for each class $c$ has been built, the adversary trains a shadow model $A$ to mimic model $S$. 
As the adversary knows which transfer learning method is used to train model $S$, 
she, thus, adopts the same method to train the shadow model $A$ using a cross-entropy loss function.

\subsubsection{Step 2. Attack Model Training}
After training shadow model $A$, the adversary trains an attack model $G$ to invert model $A$. 
The training of model $G$ is based on the regular training process \cite{Yang19}. 
That is, shadow model $A$ takes a sample $x$ as input and produces a prediction vector $A(x)$ 
that is then taken as an input by model $G$. 
The reconstructed data is then output as $\hat{x}=G(A(x))$. 
The training samples of model $G$ are the same as the shadow model $A$, 
and the loss function is mean squared error. 

\subsubsection{Step 3. Student Model Attacking}
Once model $G$ is trained, it is used to invert the student model $S$.  
The inversion process is similar to the first method. 

\section{Discussion of the Attack Methods}
\subsection{Discussion of the First Attack Method}\label{sub:discussion1}
In the first method, the adversary uses both the teacher data and student data to train the student inversion model. 
The reasoning is explained as follows. 

Formally, the loss function used to train both the teacher and student inversion models is mean squared error: 
\begin{equation}\label{eq:loss}
    L=\frac{1}{m}\sum^m_{i=1}||G(F(x_i))-x_i||^2_2, 
\end{equation}
where $m$ is the number of samples used to train an inversion model. 
Ideally, Eq. \ref{eq:loss} can be rewritten as: 
\begin{equation}
    L=\int_{\mathbb{R}^d}p(x)||G(F(x))-x||^2_2dx,
\end{equation}
where $\mathbb{R}^d$ denotes the data space of training samples and $d$ means the dimension of each sample. 
Let $f=p(x)||G(F(x))-x||^2_2$ and $r(x)=G(F(x))$. We have 
\begin{equation}
    f=p(x)||r(x)-x||^2_2. 
\end{equation}
Then, according to the Euler-Lagrange equation \cite{Dacorogna04,Alain14}, to be satisfied at the optimal $r$, we have 
\begin{equation}
    \frac{\partial f}{\partial r}-\frac{\partial}{\partial x}(\frac{\partial f}{\partial r'})=0,
\end{equation}
where $r'=\frac{\partial r}{\partial x}$. 
Since $f$ is independent of $r'$, then $\frac{\partial}{\partial x}(\frac{\partial f}{\partial r'})=0$. 
Hence, we have 
\begin{equation}
    \frac{\partial f}{\partial r}=2p(x)(r(x)-x)=0.
\end{equation}
This means that to achieve the optimal $r$, $r(x)$ must be equal to $x$, given that $p(x)\neq 0$. 
However, this equation is hard to achieve in reality. 
Thus, the data distribution $p(x)$ plays an important part in the performance of $r$. 

As the adversary only has a limited amount of student data, 
the student data distribution may not be known. 
Therefore, if the adversary uses only these student data to train a student inversion model, its performance may not be satisfactory. 
To address this issue, the teacher data are incorporated into the training of the student inversion model. 
Given that the teacher data and the student data have a similar distribution, 
involving the teacher data can reveal a proper distribution 
which improves the performance of the student inversion model, as shown in the experiments. 

\subsection{Discussion of the Second Attack Method}\label{sub:discuss 2nd method}
The key step in the second method is augmenting the student data using the noise-based and the jigsaw-based approaches. 
As summarized in \cite{Shorten19}, there are many image data augmentation approaches, 
such as flipping, rotating, and editing images directly \cite{Wang18CVPR,Chen19TIP}. 
These approaches can improve a model' classification accuracy. 
However, it is not clear whether they can improve the reconstruction quality of inversion models. 
Generally, improving a model's classification accuracy does not automatically improve the reconstruction quality of an inversion model 
that has been trained against that classification model. 
This is because training these two types of models has different goals. 
To train a classification model well, one must improve its generalization ability, 
while, to train an inversion model well, one must let it see as diverse a selection of samples as possible. 
Specifically, when training a classification model, the loss function usually compares the difference 
between predicted labels and true labels of training samples. 
When training an inversion model, the loss function compare the difference 
between reconstructed samples and real samples. 
Thus, a classification model is trained to distinguish different labels, 
while an inversion model is trained to distinguish different samples. 
Hence, for a given sample, the adversary has to generate both the neighboring and large-distanced samples.

With the noise-based approach, in Eq. \ref{eq:mask}, the goal of training a mask is to compactly delete image regions that are not class-specific. 
Namely, deleting these regions from an image does not affect the classification of this image. 
Since the class-specific features of a given image are preserved, 
all the images generated by the noise-based approach have the same class-specific features as the given image. 
Hence, these images are neighbors of the given image. 

\begin{figure}[ht]
\centering
	\includegraphics[scale=0.6]{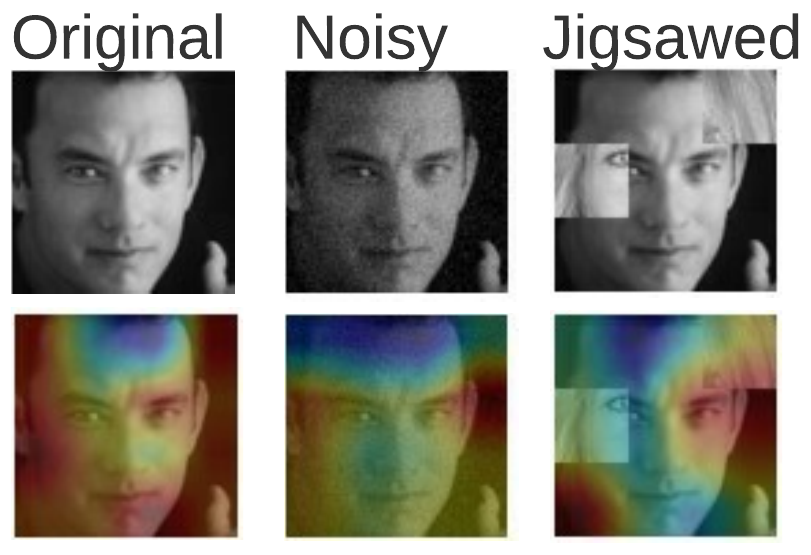}
	\caption{Class activation mapping (CAM) \cite{Zhou16} visualizations on original, noisy and jigsawed samples. Note that the noisy sample shares the same main class-specific regions as the original sample, while the jigsawed sample has different class-specific regions to the original sample.} 
	\label{fig:visualization}
\end{figure}

With the jigsaw-based approach, the goal is to generate the images that are far away but share the same main features as the given image. 
To achieve this goal, the adversary defines a learning problem, Eq. \ref{eq:mask2}, 
which aims 1) to minimize the difference in the teacher model's output at the $(N-K)$th layer between a given image and its jigsawed version; 
and 2) not to minimize the use of the contents from auxiliary samples. 
For the first aim, the teacher model is simply used as a feature extractor. 
Given that the teacher model has been well-trained,  
it can extract the features of images of unseen classes. 
This claim is supported by the self-supervised learning research \cite{Noroozi18,Jing21,Liu21}, 
where researchers often use well-trained image classifiers, such as VGG16 \cite{Simonyan15}, to extract the features of unlabeled images for unseen classes. 
For the second aim, the term $\lambda\cdot||m_c||_1$ is removed to avoid masking a major part of the auxiliary samples. 
Figure \ref{fig:visualization} visualizes the class-specific image regions. 
We can see that the noisy sample has similar class-specific regions to the original one, 
while the jigsawed sample lets a model focus on different regions to the original one. 
The visualization results prove our hypothesis. 

Next, we need to discuss the upper bound of inversion quality reduction between using synthetic data and genuine data. 
Let the reconstruction result be $r(x)=G(F(x))$, and assume that $r(x)$ is $l$-Lipschitz continuous, 
i.e., for any $x_1$, $x_2$ in the sample space, $||r(x_1)-r(x_2)||_2\leq l||x_1-x_2||_2$. 
Lipschitz continuity in a network can be achieved in many ways: by adding a penalty term to the loss function \cite{Kim20, Barrett21} 
or by performing a constrained optimization when training the network to constrain the Lipschitz constant of each layer \cite{Gouk21}. 

Given a genuine sample $x$ and its synthetic version $x+\eta$, the reconstructed results are $r(x)$ and $r(x+\eta)$, respectively. 
Since $r$ is $l$-Lipschitz continuous, we have: $||r(x+\eta)-r(x)||_2\leq l||\eta||_2$. 
According to the loss function in Eq. \ref{eq:loss}, the inversion quality of a given sample $x$ is measured by $||r(x)-x||_2$. 
Hence, the inversion quality of a synthetic sample $x+\eta$ is $||r(x+\eta)-(x+\eta)||_2$. 
Then, the upper bound of the inversion quality difference between a genuine sample $x$ and a synthetic sample $x+\eta$ can be computed as 
\begin{equation}\label{eq:bound}
\begin{aligned}
    &||r(x+\eta)-(x+\eta)||_2-||r(x)-x||_2\\
    &\leq||r(x+\eta)-(x+\eta)-(r(x)-x)||_2\\
    &=||r(x+\eta)-r(x)+(x-(x+\eta))||_2\\
    &\leq||r(x+\eta)-r(x)||_2+||\eta||_2\\
    &\leq l||\eta||_2+||\eta||_2=(l+1)||\eta||_2, 
\end{aligned}
\end{equation}
where the first and second inequality is based on the subadditivity property of norms, 
and the third inequality comes from the definition of Lipschitz continuity. 
As Lipschitz constant $l>0$, the bound $(l+1)||\eta||_2>0$. 
This result demonstrates that the inversion quality of using synthetic data is inevitably worse than using genuine data. 
This case will be shown experimentally in the next section. 

\section{Experiments}\label{sec:experiments}
We evaluated our methods by comparing it with an existing work on four datasets. 
Two factors were evaluated in the experiments: the number of student model classes and the student data size. 
We also estimate two numerical metrics: the data inversion error and the confidence vector error.

\subsection{Experimental Setup}
\subsubsection{Datasets}
The four datasets we chose for the experiments have been broadly used in related studies. 

\noindent\textbf{FaceScrub} \cite{Ng14} - a dataset of URLs for $100,000$ images of 530 individuals. 
    We collected $91,712$ images of $526$ individuals.
    Each image was resized to $64\times 64$.
    
\noindent\textbf{CelebA} \cite{Liu15} - a dataset with $202,599$ images of $10,177$ celebrities, i.e., classes, from the Internet. 
    After removing $296$ celebrities that are also included in FaceScrub, we were left with $195,727$ images of $9,881$ celebrities. 
    Again, each image was resized to $64\times 64$. 

\noindent\textbf{MNIST} \cite{LeCun98} - a dataset of $70,000$ images of handwritten numerals spanning $10$ classes: $0-9$. 
Each image was resized to $32\times 32$.
    
\noindent\textbf{Fashion-MNIST} \cite{Fashion} consists of $70,000$ images across $10$ classes, 
    including T-shirt, trouser, pullover, dress, coat, sandal, shirt, sneaker, bag and ankle boot. 
    Again, each image was resized to $32\times 32$. 
    

\subsubsection{Model Training}    
\begin{table}[ht]\scriptsize
	\caption{Performance of different tasks using different transfer learning methods. 
	With the deep-layer feature extractor, we froze the $N_T-1$ layers of the teacher model and only updated the last classification layer. 
	With the mid-layer feature extractor, we froze the CNN layers of the teacher model while the succeeding fully-connected layers were updated.}
\centering
\small
\begin{tabular}{|c|c|c|c|} \hline
Student task & Deep-layer & Mid-layer & Full-model\\ \hline
FaceScrub & \textbf{95.32\%} & 94.63\% & 93.28\% \\ \hline
MNIST & 93.07\% & 93.38\% & \textbf{94.11\%} \\ \hline
\end{tabular}
	\label{tab:training}
\end{table}

\noindent\textbf{The first attack method} \hspace{2mm} 
There are six models: the teacher model $T$, the modified teacher model $T_E$, the shadow model $A$, 
the attack model $G$, the student model $S$ and the conversion model $I$. 
Among them, model $T_E$ is directly inherited from the teacher model $T$ and does not need training. 
Moreover, the shadow model $A$, transferred from model $T$, is connected with the conversion model $I$. 
Hence, models $A$ and $I$ are trained together. 

We divided FaceScrub dataset into two sets. 
The first set with $426$ classes was used to build the teacher model $T$. 
The second set with $100$ classes was used to build the student model $S$. 
For models $T$ and $S$, $80\%$ of the data were used for training and $20\%$ for testing. 
To optimize model $S$, we have tried the three transfer learning methods: 
deep-layer feature extractor, mid-layer feature extractor, and full model fine-tuning. 
The results are shown in Table \ref{tab:training}. It can be seen that the deep-layer feature extractor was the best method. 

The attack model $G$ was then used to invert model $T$ and hence, was trained with the same set as model $T$. 
The shadow model $A$ and conversion model $I$ were connected together and trained using the teacher data $D_T$ and student data $D_S$. 
Here, we set $D_T$ to be the same as the training set of model $T$, 
randomly selecting $10$ samples for each class from the test set of model $S$ to form $D_S$. 

With CelebA, as each class has only about $20$ samples, it cannot be used to properly train a teacher model. 
Thus, it was only used as an auxiliary set in the second attack method. 



With MNIST and Fashion-MNIST, we used Fashion-MNIST to train the teacher model $T$ and MNIST to train the student model $S$. 
To imitate a transfer learning scenario, we randomly selected $100$ samples for each class from MNIST to train the student model. 
We used the same three transfer learning methods to optimize the student model. 
The results are given in Table \ref{tab:training}. 
As the two datasets share few common features, full model fine-tuning was the best method. 
Likewise, we set $D_T$ to be the same as the training set of model $T$, 
randomly selecting $10$ samples for each class from the test set of model $S$ to form $D_S$.

\noindent\textbf{The second attack method} \hspace{2mm} 
As the adversary does not have any teacher data now, 
she is not capable of training an attack model to invert the teacher model. 
Thus, there are only four models: the teacher model $T$, the student model $S$, the shadow model $A$, and the attack model $G$. 
Models $T$ and $S$ were trained in the same way as the first method. 
For the shadow model $A$ and the attack model $G$, the training data were the student data $D_S$ plus the augmented data $D_{aug}$. 

We divided FaceScrub into two sets. 
The first set with $426$ classes was used to build the teacher model $T$. 
The second set with $100$ classes was used to build the student model $S$. 
We used CelebA as the auxiliary set to augment the student data $D_S$ to form $D_{aug}$. 
For each sample $x$ in $D_S$, we use the noise-based augmentation approach to create $20$ samples. 
Then, we randomly select $20$ samples, based on the gender of the person in $x$, from CelebA, 
and used the jigsaw-based augmentation approach to synthesize another $20$ samples. 

For MNIST and Fashion-MNIST, models $T$ and $S$ were trained in the same way as the first method. 
However, in this case, the adversary has no teacher data. 
Thus, to augment the student data $D_S$, 
we first used the noise-based augmentation method on each sample $x$ in $D_S$ to synthesize $20$ samples. 
Upon applying the jigsaw-based augmentation method to the MNIST samples, we found it was not working very well. 
This may be because the contents of the MNIST samples are simple. 
Thus, cutting and replacing even a small piece of an image may destroy its main features. 
Hence, we used the standard rotation and slipping approach instead to create another $20$ samples for each sample $x$ in $D_S$. 

\noindent\textbf{Discussion} \hspace{2mm} Another popular method used in transfer learning experiments 
is using already established models as teacher models. 
For example, the VGG-face model \cite{Simonyan15}, trained on a dataset of $2.6$M images to recognize $2,622$ faces, 
can be used as a teacher model to train other student facial classifiers. 
We have also tried this method in our experiments. 
However, as VGG-face model has 16 layers, it was hard to invert using our equipment. 

\subsubsection{Model Architecture}
We used the model architecture proposed in \cite{Yang19}. 
The FaceScrub teacher model $T$ included 4 CNN blocks, two fully-connected layers and a softmax function. 
Each CNN block consisted of a convolutional layer followed by a batch normalization layer, 
a max-pooling layer and a ReLU activation layer. 
The two fully-connected layers were added after the CNN blocks. 
The softmax function was added to the last layer to convert neural signals into a valid confidence vector $\mathbf{y}$, 
where each $y_i\in [0,1]$ and $\sum^{k}_{i=1} y_i=1$, and $k$ is the number of classes. 
The FaceScrub student model $S$ has the same architecture as the teacher model. 
When the student model used deep-layer feature extractor, 
the 4 CNN blocks and a fully-connected layer were frozen, while the last fully-connected layer was updated. 
When using the mid-layer feature extractor, the 4 CNN blocks were frozen, while the two fully-connected layers were updated. 
When using full model fine-tuning, the whole student model was updated. 
Similar to the FaceScrub configuration, the Fashion-MNIST teacher model and MNIST student model had 3 CNN blocks. 

The attack model $G$ against the FaceScrub teacher model $T$ included 5 transposed CNN blocks. 
Each of the first 4 blocks had a transposed convolutional layer followed by a batch normalization layer and a $Tanh$ activation function. 
The last block had a transposed convolutional layer succeeded by a Sigmoid activation function that converted neural signals into real values in $[0,1]$. 
Akin to the FaceScrub configuration, the attack model against the MNIST student model had 4 transposed CNN blocks. 

The shadow model $A$ used to mimic the student model $S$ had the same architecture as the student model 
but they have different number of neurons in the fully-connected layers. 
Note that the adversary does not know the student model's architecture, but she does know the teacher model's architecture. 
Hence, the adversary can treat the shadow model's architecture as a hyper-parameter and tune it via experiments. 

In the first attack method, the adversary has additional two models: the modified teacher model $T_E$ and the conversion model $I$.
The architecture of model $T_E$ is similar to the teacher model $T$ by removing the fully-connected layers from model $T$. 
The conversion model $I$ is a standard three-fully-connected-layer neural network. 

\subsubsection{Comparison Attack Methods}
We compared our two methods, denoted as \emph{inversion with $D_T$} and \emph{inversion without $D_T$}, with the \emph{direction inversion} method from \cite{Yang19}. 

\noindent\textbf{Direct inversion} \cite{Yang19}: An adversary trains an attack model using the outputs of the target model by directly querying it. 
To achieve the best performance with the \emph{direct inversion} method, 
the adversary is given $50\%$ amount of the target model's testing set which is used to train the attack model. 

The \emph{direct inversion} method is a standard model inversion method in the literature. 
Although new methods have been proposed recently \cite{Zhao21,Salem20,Carlini21}, 
as discussed in Section \ref{sec:related work}, they were developed for settings different to ours. 


\subsubsection{Evaluation Factors}
\noindent\textbf{The number of classes of the student model}: 
This is the number of classes that the student model can classify. 
This factor simulates the situations, where student models have different classification tasks. 

\noindent\textbf{The size of the student data}: 
This is the number of samples possessed by the adversary. 
It simulates the situations, where adversaries have different capability to collect data. 

\subsubsection{Evaluation Metrics}
\noindent\textbf{Data inversion error} shows how accurate the attack model is at reconstructing the input data. 
It is measured by computing the mean squared error between an input data record and its reconstructed version. 
Note that any input data records to the target student model are  unseen to the adversary. 
Here, we use them only for evaluating the experimental results.

\noindent\textbf{Confidence vector error} shows the accuracy of the shadow model in predicting the confidence vectors. 
It is measured by computing the Euclidian distance between a real confidence vector output by the student model 
and the predicted confidence vector output by the shadow model. 
Again, in the real world, the input data to the student model cannot be accessed by the adversary. 
Thus, the real confidence vectors output by the student model cannot be obtained by the adversary. 
They were only used to evaluate the experimental results. 

\subsection{Experimental Results}
\subsubsection{General Inversion Quality}
\begin{figure}[ht]
\centering
	\includegraphics[scale=0.49]{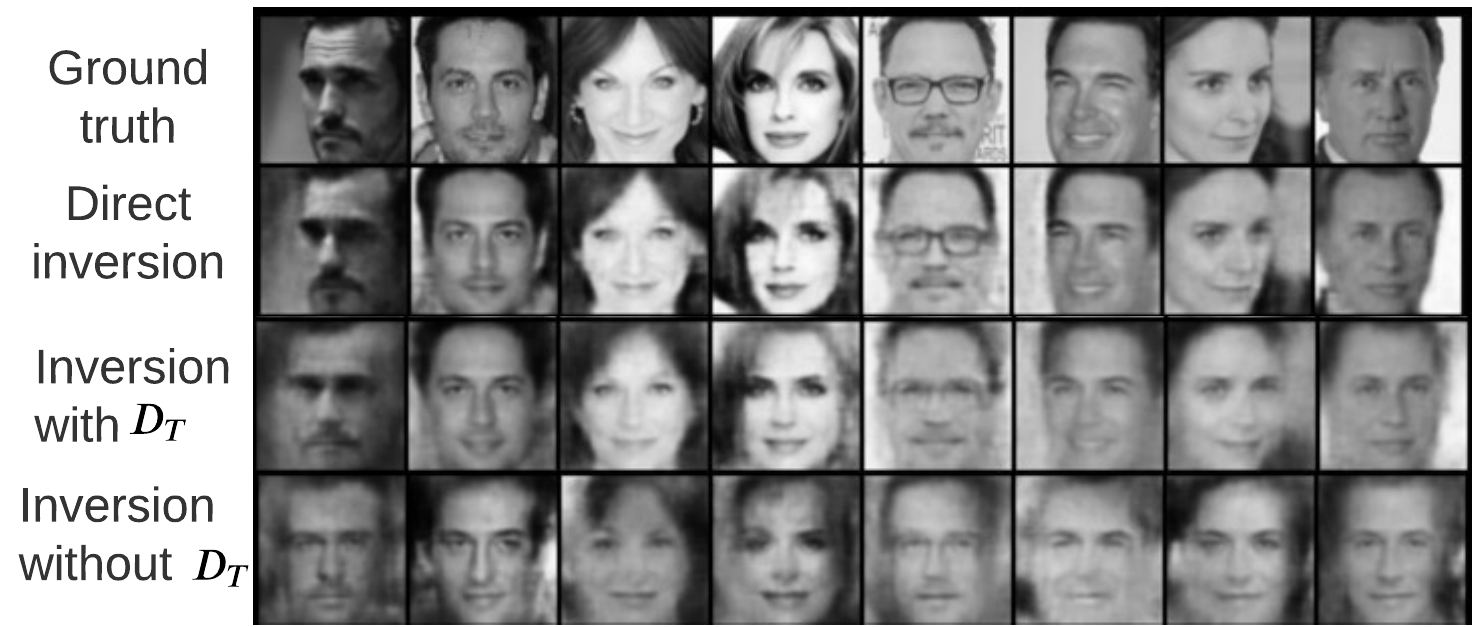}
	\caption{Performance of the three methods on FaceScrub} 
	\label{fig:FaceScrub}
\end{figure}

\begin{figure}[ht]
\centering
	\includegraphics[scale=0.32]{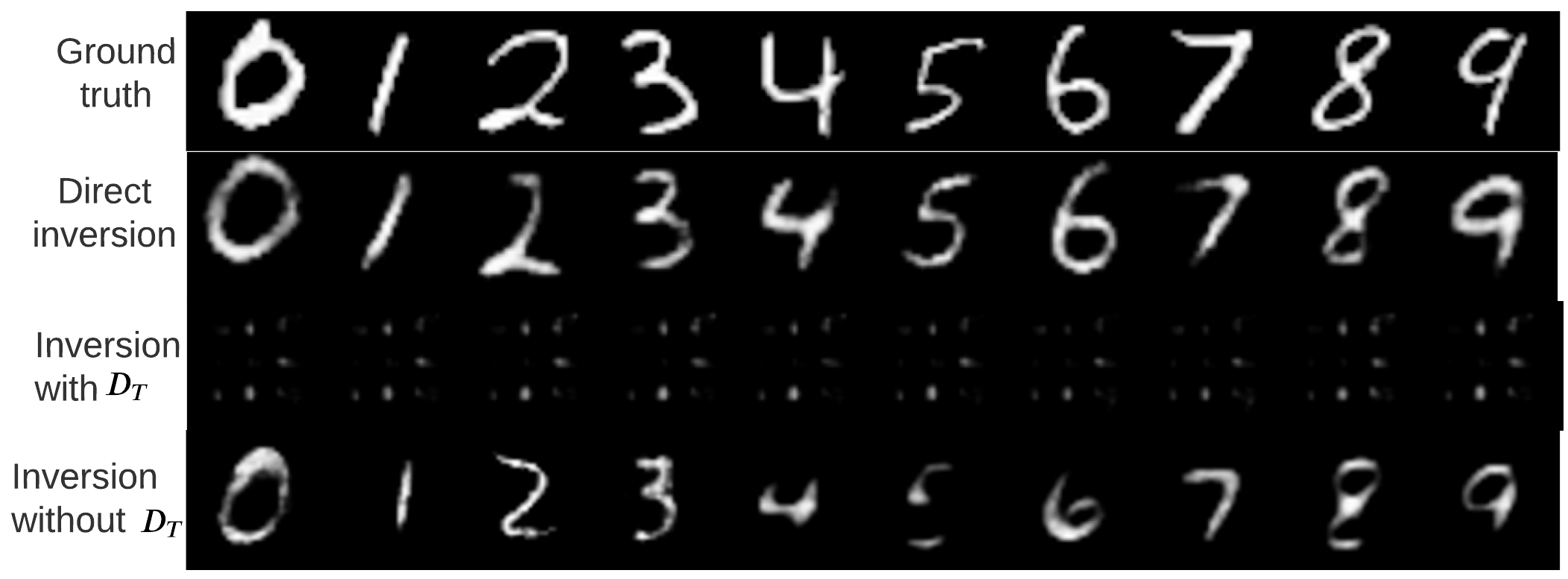}
	\caption{Performance of the three methods on MNIST} 
	\label{fig:MNIST}
\end{figure}

Figures \ref{fig:FaceScrub} and \ref{fig:MNIST} show the inversion quality of the three methods on FaceScrub and MNIST, respectively. 
In Figure \ref{fig:FaceScrub}, our first method, \emph{inversion with $D_T$}, achieves almost the same results as the \emph{direct inversion} method. 
Although the inversion quality of our second method, \emph{inversion without $D_T$}, is slightly worse than the \emph{direct inversion} method, 
the reconstructed images are still recognizable to humans. 
This performance difference is mainly incurred by the inaccessibility of querying the student model in our methods. 
In Figure \ref{fig:MNIST}, our first method, \emph{inversion with $D_T$}, did not work well. 
This may be because the teacher data $D_T$ is from Fashion-MNIST which shares few common features with MNIST. 
As analyzed in Section \ref{sub:discussion1}, using these teacher data may affect the distribution of student data, 
which impairs the performance of the attack model. 
However, we can still see that the performance of our second method, \emph{inversion without $D_T$}, 
is almost as good as the \emph{direct inversion} method. 
This result shows that our two methods suit different situations. 

The above results are also supported by the quantitative data inversion errors as shown in Table \ref{tab:Inversion}, 
where higher inversion quality corresponds a smaller inversion error. 
Moreover, in Table \ref{tab:confidence}, we can see that there is a small gap between the real confidence vectors and the predicted confidence vectors. 
This gap explains why the inversion quality of our methods is slightly worse than the \emph{direct inversion} method. 
\vspace{-3mm}
\begin{table}[!ht]\scriptsize
\caption{Data inversion errors of the three methods}
	\centering
\begin{tabular}{|c|c|c|} \hline
& FaceScrub & MNIST \\ \hline
Direct inversion & $0.834$ & $0.821$ \\ \hline
Inversion with $D_T$ & $0.857$ & $0.863$ \\ \hline
Inversion without $D_T$ & $0.872$ & $0.835$ \\ \hline
\end{tabular}
	\label{tab:Inversion}
\end{table}
\vspace{-4mm}
\begin{table}[!ht]\scriptsize
\caption{Confidence vector errors of our methods}
	\centering
\begin{tabular}{|c|c|c|} \hline
& FaceScrub & MNIST \\ \hline
Inversion with $D_T$ & $0.042$ & $0.033$  \\ \hline
Inversion without $D_T$ & $0.053$ & $0.021$ \\ \hline
\end{tabular}
	\label{tab:confidence}
\end{table}
\vspace{-3mm}
\subsubsection{Effects of the Number of Student Model Classes}
\begin{figure}[ht]
\centering
	\includegraphics[scale=0.49]{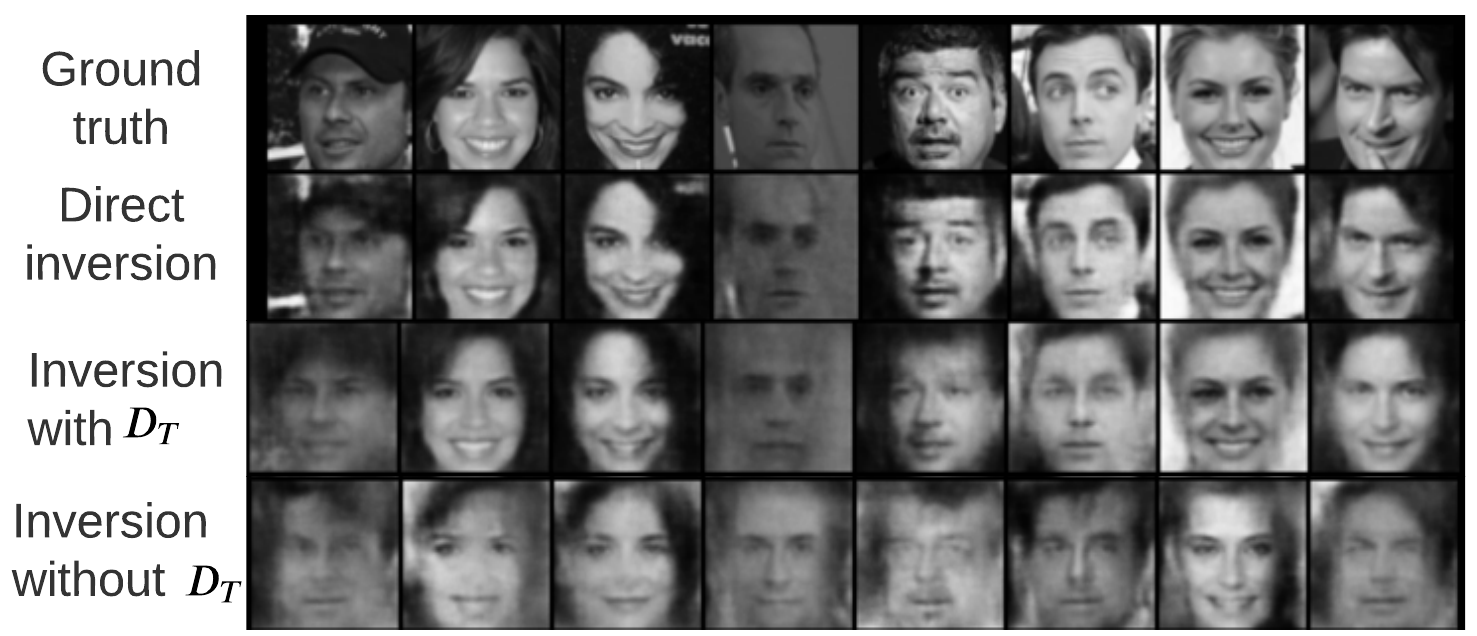}
	\caption{Performance of the three methods on FaceScrub when the number of student model classes is $50$} 
	\label{fig:FaceScrub50}
\end{figure}

\begin{figure}[ht]
\centering
	\includegraphics[scale=0.49]{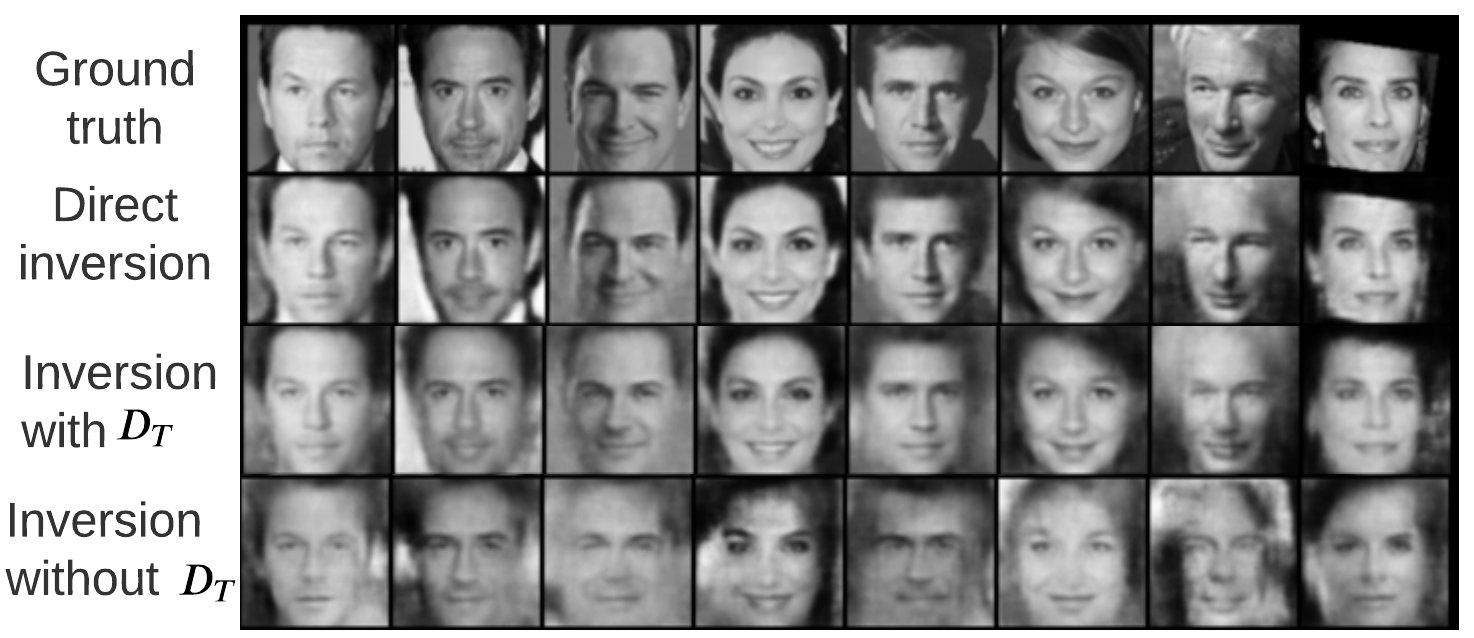}
	\caption{Performance of the three methods on FaceScrub when the number of student model classes is $100$} 
	\label{fig:FaceScrub100}
\end{figure}

\begin{figure}[ht]
\centering
	\includegraphics[scale=0.33]{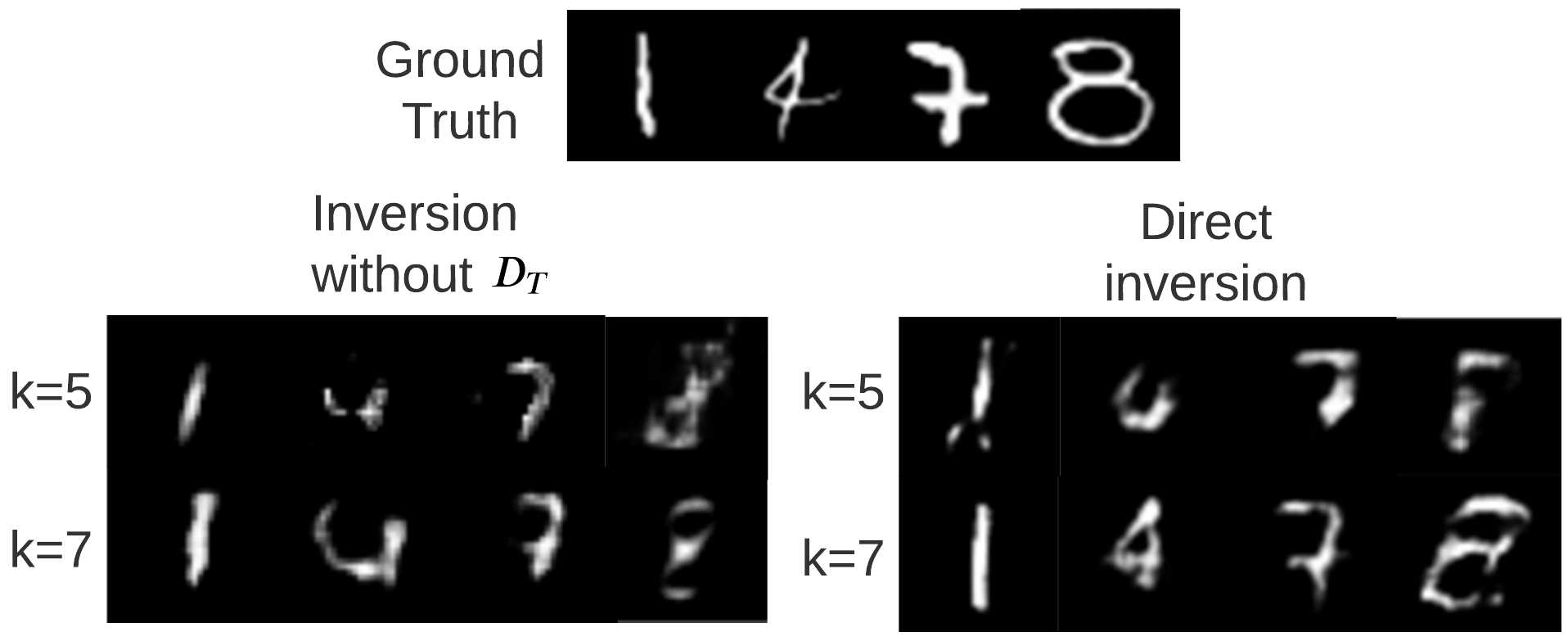}
	\caption{Performance of the two methods on MNIST with different number of student model classes} 
	\label{fig:MNISTClass}
\end{figure}

In the remaining experiments, the results for \emph{inversion with $D_T$} are not given, 
as it did not perform well on MNIST. 

Figures \ref{fig:FaceScrub50}, \ref{fig:FaceScrub100} and \ref{fig:MNISTClass} show the inversion quality of the three methods on FaceScrub and MNIST with different numbers of classes. 
The overall results demonstrate that the inversion quality of the three methods improves as the number of classes increases. 
This can be explained by the fact that in such model inversion attacks, 
the target model, or the shadow models in our methods, is used as an encoder, 
whose output contains the information about the input sample. 
As a high-dimension output embodies more information than a low-dimension output, 
the student model with more output classes lead to better inversion quality. 
In addition, in Tables \ref{tab:inversion error class} and \ref{tab:confidence error class}, the quantitative measurement also shows that 
the data inversion error and confidence vector error are inversely proportional to the number of classes. 
\vspace{-3mm}
\begin{table}[!ht]\scriptsize
\caption{Data inversion errors of the three methods on FaceScrub and MNIST with different number of student classes}
	\centering
\begin{tabular}{|c|c|c|c|} \hline
& & MNIST & FaceScrub \\ \hline
\multirow{4}{*}{Direct inversion} & $5$ classes & $0.863$ & ---- \\ \cline{2-4}
& $7$ classes & $0.845$ & ---- \\ \cline{2-4}
& $50$ classes & ---- & $0.873$ \\ \cline{2-4}
& $100$ classes & ---- & $0.855$ \\ \hline
\multirow{2}{*}{Inversion with $D_T$} 
& $50$ classes & ---- & $0.871$ \\ \cline{2-4}
& $100$ classes & ---- & $0.858$ \\ \hline
\multirow{4}{*}{Inversion without $D_T$} & $5$ classes & $0.895$ & ---- \\ \cline{2-4}
& $7$ classes & $0.887$ & ---- \\ \cline{2-4}
& $50$ classes & ---- & $0.886$ \\ \cline{2-4}
& $100$ classes & ---- & $0.874$ \\ \hline
\end{tabular}
	\label{tab:inversion error class}
\end{table}
\vspace{-4mm}
\begin{table}[!ht]\scriptsize
\caption{Confidence vector errors of our methods on FaceScrub and MNIST with different number of student classes}
	\centering
\begin{tabular}{|c|c|c|c|} \hline
& & MNIST & FaceScrub \\ \hline
\multirow{2}{*}{Inversion with $D_T$} 
& $50$ classes & ---- & $0.048$ \\ \cline{2-4}
& $100$ classes & ---- & $0.043$ \\ \hline
\multirow{4}{*}{Inversion without $D_T$} & $5$ classes & $0.039$ & ---- \\ \cline{2-4}
& $7$ classes & $0.031$ & ---- \\ \cline{2-4}
& $50$ classes & ---- & $0.062$ \\ \cline{2-4}
& $100$ classes & ---- & $0.057$ \\ \hline
\end{tabular}
	\label{tab:confidence error class}
\end{table}
\vspace{-3mm}
\subsubsection{Effects of the Student Data Size}\label{sub:data size}
\begin{figure}[ht]
\centering
	\includegraphics[scale=0.49]{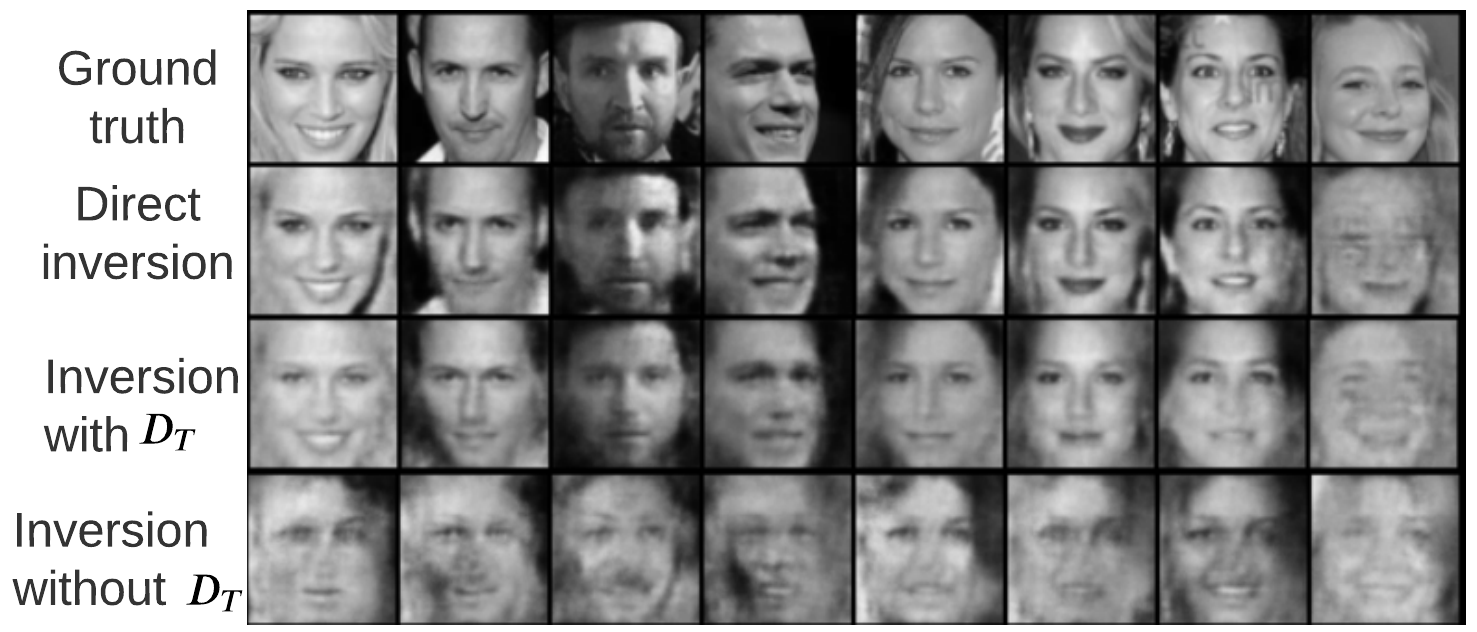}
	\caption{Performance of the three methods on FaceScrub when the number of student data in each class is $7$} 
	\label{fig:FaceScrub7}
\end{figure}

\begin{figure}[ht]
\centering
	\includegraphics[scale=0.49]{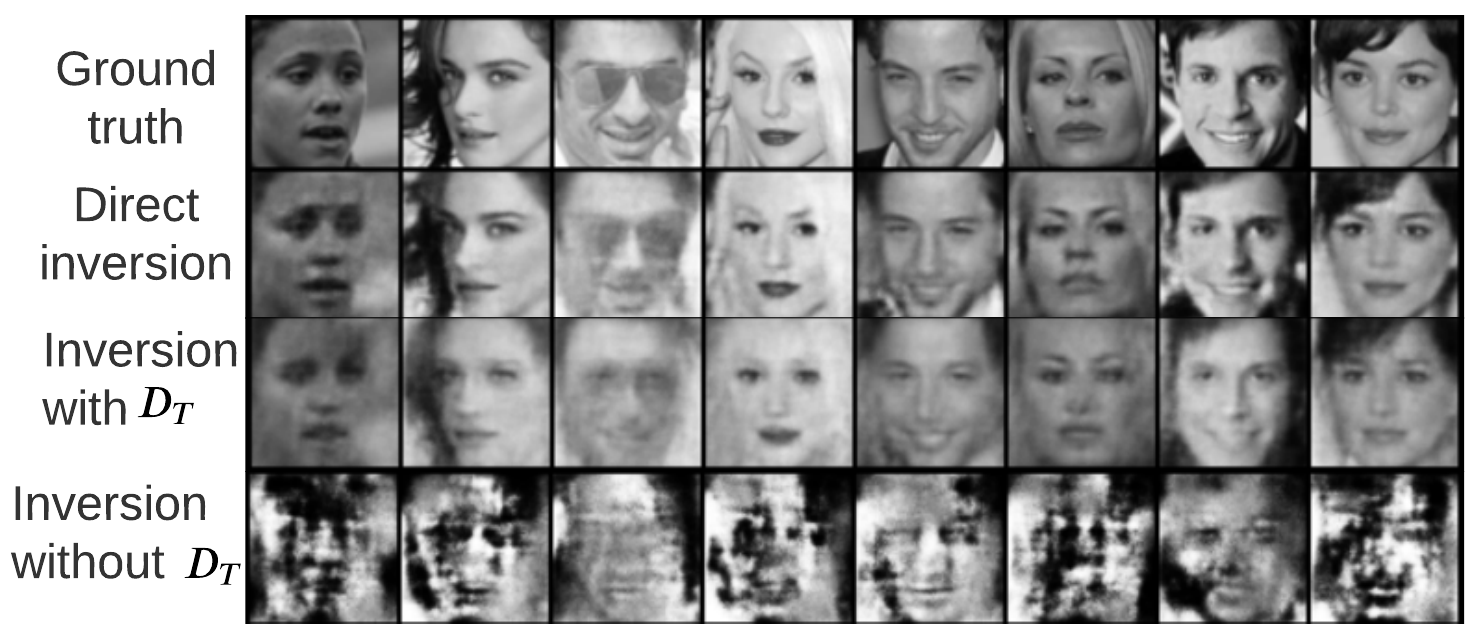}
	\caption{Performance of the three methods on FaceScrub when the number of student data in each class is $3$} 
	\label{fig:FaceScrub3}
\end{figure}

\begin{figure}[ht]
\centering
	\includegraphics[scale=0.39]{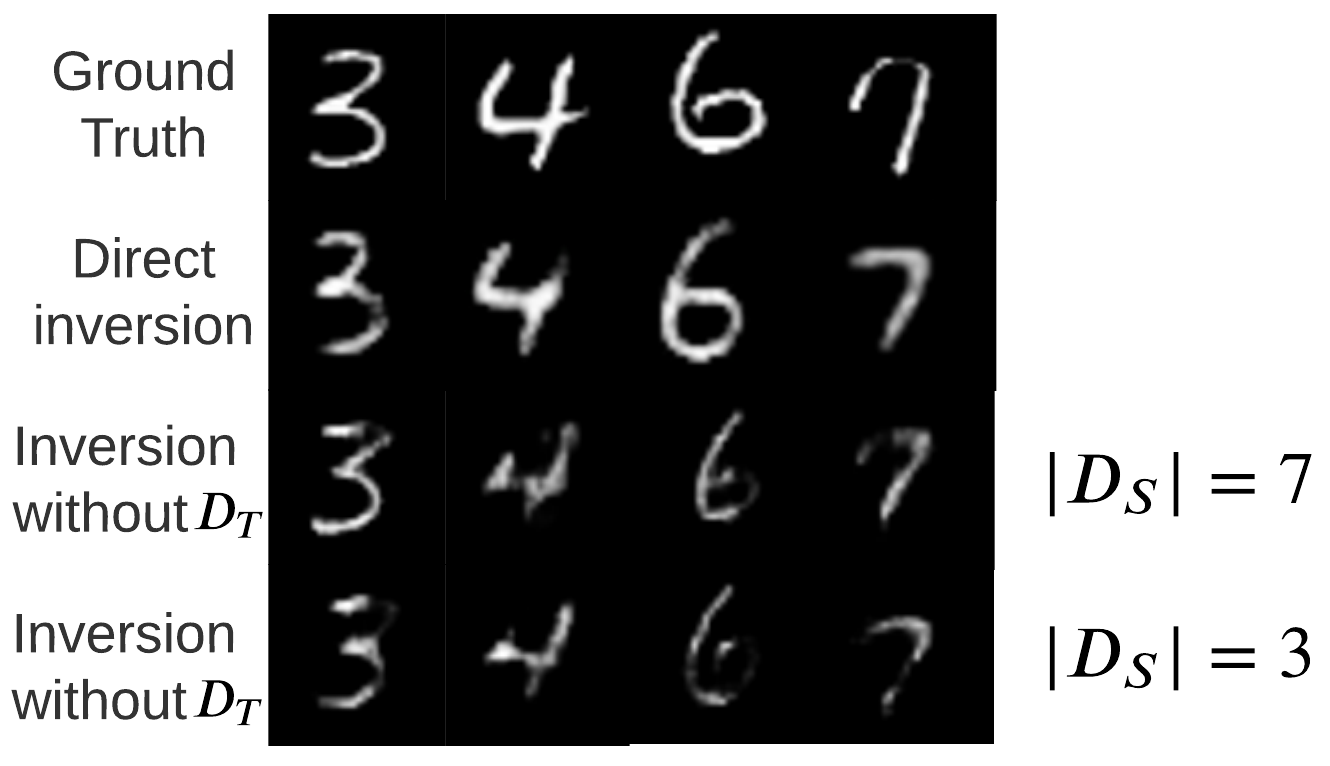}
	\caption{Performance of the two methods on MNIST with different amounts of student data} 
	\label{fig:MNISTData}
\end{figure}

Figures \ref{fig:FaceScrub7}, \ref{fig:FaceScrub3} and \ref{fig:MNISTData} show the inversion quality 
with different-sized student data for our methods on FaceScrub and MNIST. 
Note that the \emph{direct inversion} method still uses a massive amount of student data to guarantee good performance. 
It can be seen that the student data size has impact on the inversion quality of our methods on FaceScrub but has little effects on MNIST. 
This may be because facial images are much more complex than hand-written digits. 
Therefore, when there are too few images in the student model's training set, 
it fails to properly train an attack model that can capture the distribution of student data. 
By comparison, as the hand-written digits are less complex, 
even a few samples from the student data were enough to train an attack model to capture the distribution of student data. 
This result is also shown by the quantitative measurements in Tables \ref{tab:inversion error data} and \ref{tab:confidence error data}. 
\vspace{-3mm}
\begin{table}[!ht]\scriptsize
\caption{Data inversion errors of the three methods on FaceScrub and MNIST with different student data size}
	\centering
\begin{tabular}{|c|c|c|c|} \hline
& & MNIST & FaceScrub \\ \hline
\multirow{2}{*}{Direct inversion} & $3$ samples & $0.913$ & $0.935$ \\ \cline{2-4}
& $7$ samples & $0.889$ & $0.901$ \\ \hline
\multirow{2}{*}{Inversion with $D_T$} & $3$ samples & ---- & $0.942$ \\ \cline{2-4}
& $7$ samples & ---- & $0.917$ \\ \hline
\multirow{2}{*}{Inversion without $D_T$} & $3$ samples & $0.923$ & $0.946$ \\ \cline{2-4}
& $7$ samples & $0.897$ & $0.922$ \\ \hline
\end{tabular}
	\label{tab:inversion error data}
\end{table}
\vspace{-4mm}
\begin{table}[!ht]\scriptsize
\caption{Confidence vector errors of our methods on FaceScrub and MNIST with different student data size}
	\centering
\begin{tabular}{|c|c|c|c|} \hline
& & MNIST & FaceScrub \\ \hline
\multirow{2}{*}{Inversion with $D_T$} & $3$ samples & ---- & $0.054$ \\ \cline{2-4}
& $7$ samples & ---- & $0.049$ \\ \hline
\multirow{2}{*}{Inversion without $D_T$} & $3$ samples & $0.043$ & $0.063$ \\ \cline{2-4}
& $7$ samples & $0.038$ & $0.058$ \\ \hline
\end{tabular}
	\label{tab:confidence error data}
\end{table}
\vspace{-4mm}
\subsection{Summary of Experiments}
According to the experimental results, our first method delivers almost the same performance as the \emph{direct inversion} method on FaceScrub, 
but it does not perform well on MNIST. 
By contrast, our second method performs almost as well as the \emph{direct inversion} method on MNIST, 
but is less effective than our first method on FaceScrub. 
This result shows that our two methods suit different situations. 
This result also demonstrates that even if the student model is not allowed to be queried and the adversary has only a limited number of student data, 
the student model can still be inverted. 

\section{Mitigation}
Since our attack methods use the prediction vectors of the target student model, 
the most effective mitigation method is modifying these vectors. 
For example, given a prediction vector $\mathbf{y}$, each element $y_i$ can be modified as $y_i'=\frac{e^{y_i/t}}{\sum_j e^{y_j/t}}$, 
where $t$ is a parameter named temperature. 
This method has been proven to be effective in defending against both membership inference and model inversion attacks \cite{Shokri17}. 
However, directly modifying the elements in a vector may destroy its utility, 
which is unacceptable in some situations, e.g., illness diagnosis in our motivating example. 

\begin{figure}[ht]
\centering
	\includegraphics[scale=0.49]{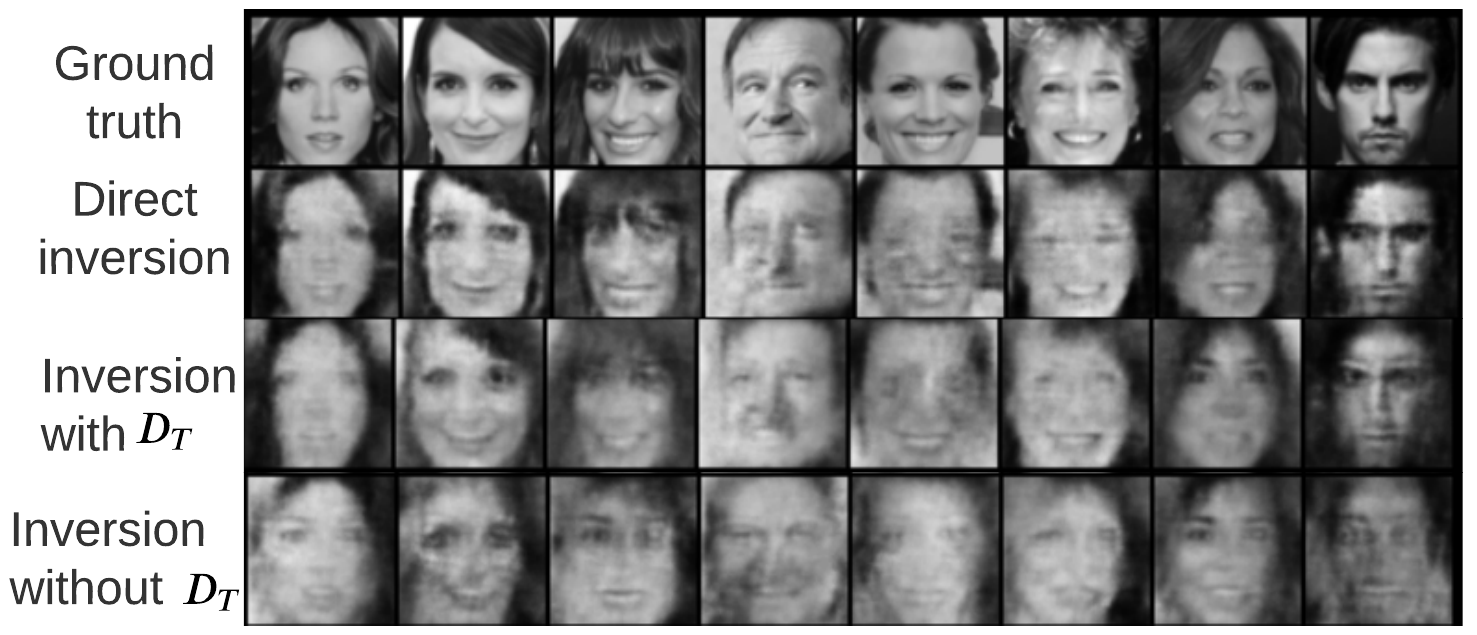}
	\caption{Defense results of the three methods on FaceScrub when the number of preserved elements, $h$, is $10$} 
	\label{fig:FaceScrubD10}
\end{figure}

\begin{figure}[ht]
\centering
	\includegraphics[scale=0.49]{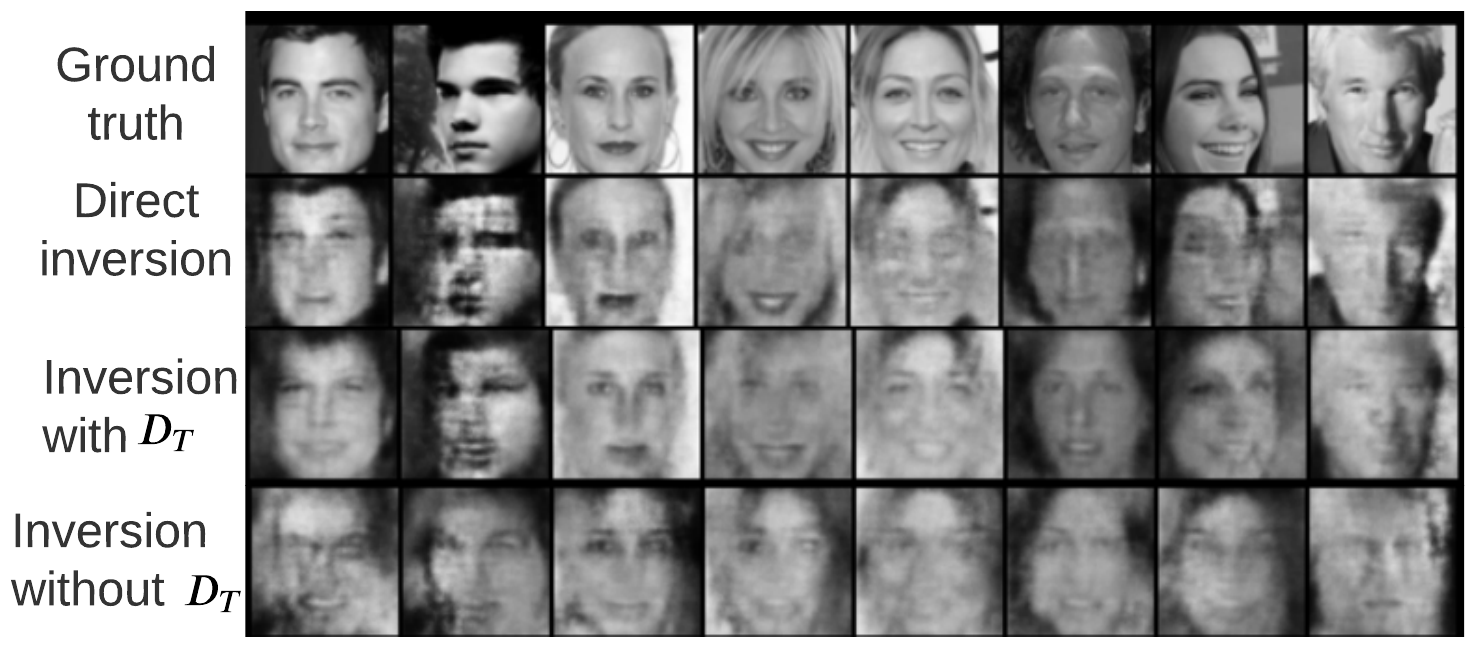}
	\caption{Defense results of the three methods on FaceScrub when the number of preserved elements, $h$, is $5$} 
	\label{fig:FaceScrubD5}
\end{figure}

\begin{figure}[ht]
\centering
	\includegraphics[scale=0.49]{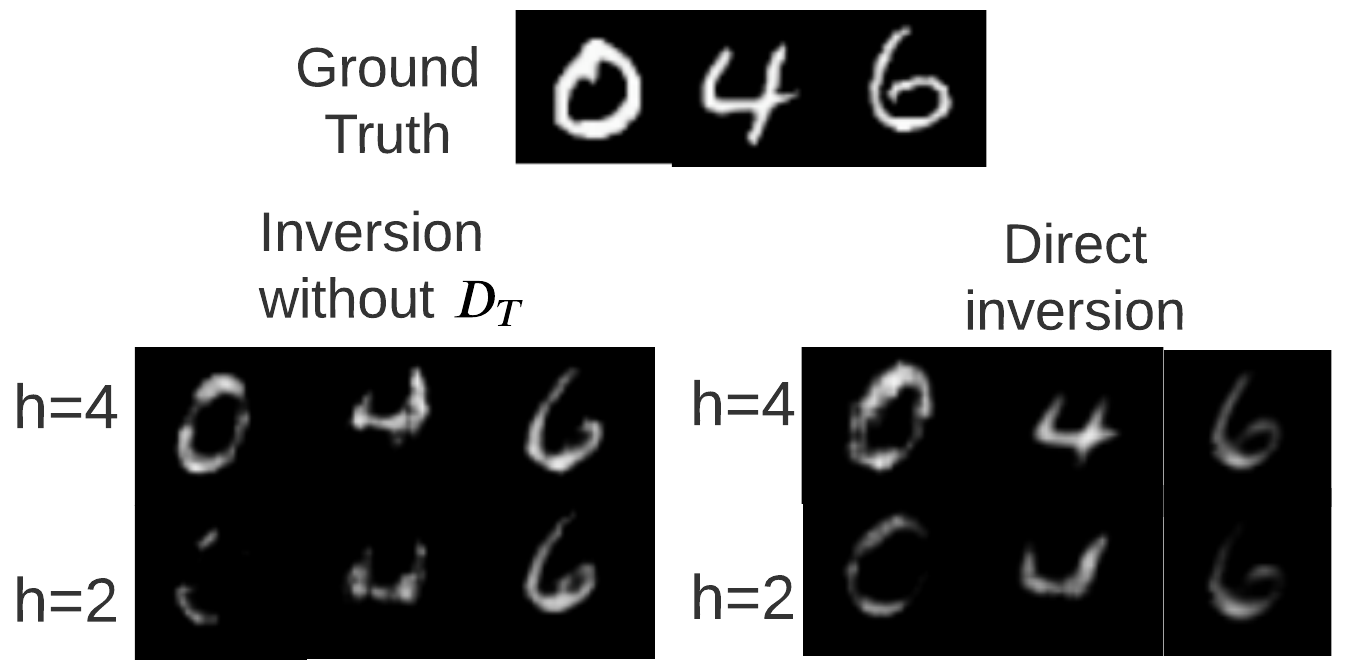}
	\caption{Defense results of the two methods on MNIST} 
	\label{fig:MNISTDefense}
\end{figure}

Thus, we adopted a restriction method to cut the prediction vector down to the top $h$ classes \cite{Shokri17}. 
The rationale is that a prediction vector typically has a few classes with large probabilities 
while the remaining classes with very small probabilities. 
Hence, the prediction vector is still useful, to some extent, if it only preserves the probabilities of the most likely $h$ classes. 
To implement this method, we added a filter to the output layer of the target student model that only preserves the largest $h$ elements. 
The smaller $h$ is, the less information the model leaks. $h$ is left as a hyper-parameter in our experiments. 
The results are shown in Figures \ref{fig:FaceScrubD10}, \ref{fig:FaceScrubD5} and \ref{fig:MNISTDefense}. 

It can be seen that this defense method does have impact on the three methods, but it needs to set a very small $h$. 
For example, for the FaceScrub classifier with $100$ classes, $h$ needs to be set to less than $10$ to achieve decent defense results. 
For the MNIST classifier, even if $h$ is set to as small as $2$, some digits can still be approximately reconstructed, e.g., `0' and `6'. 
However, a very small $h$ masks most of the information contained in output vectors, 
which may also affect the experience of benign users. 
We leave the research on appropriate defense mechanisms as our future work. 

\section{Conclusion}
In this paper, we investigated model inversion in transfer learning settings, proposing two novel and effective attack methods. 
The two methods suit different situations, depending on whether the adversary has the same-distribution data as the teacher model's training data. 
Compared with related attack methods, our methods do not require any queries to the target student model and require only a limited amount of student data. 
Our future work will focus on improving inversion quality by training a precise shadow model to mimic the target student model. 
One possible approach is to incorporate generative techniques, e.g., generative adversarial networks (GANs), to generate model parameters. 
We also plan to undertake research associated with defending against these model inversion attacks. 


\bibliographystyle{plain}
{\small \bibliography{references}}

\begin{thebibliography}{10}

\bibitem{PyTorch}
{PyTorch transfer learning tutorial}.
\newblock {\em https://pytorch.org/tutorials/beginner/transfer\_learning
  \_tutorial.html}.

\bibitem{Alain14}
G.~Alain and Y.~Bengio.
\newblock {What Regularized Auto-Encoders Learn from the Data-Generating
  Distribution}.
\newblock {\em Journal of Machine Learning Research}, (15):3743--3773, 2014.

\bibitem{Amazon}
Amazon.
\newblock {Amazon Rekognition}.
\newblock {\em https://aws.amazon.com/cn/rekognition/}, 2019.

\bibitem{Barrett21}
B.~Barrett, A.~Camuto, M.~Willetts, and T.~Rainforth.
\newblock {Certifiably Robust Variational Autoencoders}.
\newblock In {\em Proc. of NIPS}, 2021.

\bibitem{Breier21}
J.~Breier, D.~Jap, X.~Hou, S.~Bhasin, and Y.~Liu.
\newblock {SNIFF: Reverse Engineering of Neural Networks With Fault Attacks}.
\newblock {\em IEEE Transactions on Reliability}, page DOI:
  10.1109/TR.2021.3105697, 2021.

\bibitem{Carlini21}
N.~Carlini, F.~Tramer, E.~Wallace, M.~Jagielski, A.~Herbert-Voss, K.~Lee,
  A.~Roberts, T.~Brown, D.~Song, U.~Erlingsson, A.~Oprea, and C.~Raffel.
\newblock {Extracting Training Data from Large Language Models}.
\newblock In {\em Proc. of USENIX Security Symposium}, pages 2633--2650, 2021.

\bibitem{Chen21}
J.~Chen, Y.~Geng, Z.~Chen, I.~Horrocks, J.~Z. Pan, and H.~Chen.
\newblock {Knowledge-aware Zero-Shot Learning: Survey and Perspective}.
\newblock In {\em Proc. of IJCAI}, pages 4366--4373, 2021.

\bibitem{Chen19}
Z.~Chen, Y.~Fu, K.~Chen, and Y.~Jiang.
\newblock {Image Block Augmentation for One-Shot Learning}.
\newblock In {\em Proc. of AAAI}, pages 3379--3386, 2019.

\bibitem{Chen19TIP}
Z.~Chen, Y.~Fu, Y.~Zhang, Y.~Jiang, X.~Xue, and L.~Sigal.
\newblock {Multi-Level Semantic Feature Augmentation for One-Shot Learning}.
\newblock {\em IEEE Transactions on Image Processing}, 28(9):4594--4605, 2019.

\bibitem{Dacorogna04}
B.~Dacorogna.
\newblock {\em {Introduction to the Calculus of Variations}}.
\newblock World Scientific Publishing Company, 2004.

\bibitem{Fashion}
Fashion-MNIST.
\newblock {An MNIST-like dataset of 70,000 28x28 labeled fashion images}.
\newblock In {\em https://www.kaggle.com/zalando-research/fashionmnist}.

\bibitem{Fong17}
R.~C. Fong and A.~Vedaldi.
\newblock {Interpretable Explanations of Black Boxes by Meaningful
  Perturbation}.
\newblock In {\em Proc. of ICCV}, pages 3449--3457, 2017.

\bibitem{Fred15}
M.~Fredrikson, S.~Jha, and T.~Ristenpart.
\newblock {Model Inversion Attacks That Exploit Confidence Information and
  Basic Countermeasures}.
\newblock In {\em Proc. of CCS}, pages 1322--1333, 2015.

\bibitem{Fred14}
M.~Fredrikson, E.~Lantz, S.~Jha, S.~Lin, D.~Page, and T.~Ristenpart.
\newblock {Privacy in Pharmacogenetics: An End-to-End Case Study of
  Personalized Warfarin Dosing}.
\newblock In {\em Proc. of USENIX Security Symposium}, pages 17--32, 2014.

\bibitem{Geiping20}
J.~Geiping, H.~Bauermeister, H.~Droge, and M.~Moeller.
\newblock {Inverting Gradients - How easy is it to break privacy in federated
  learning?}
\newblock In {\em Proc. of NIPS}, 2020.

\bibitem{Gouk21}
H.~Gouk1, E.~Frank, B.~Pfahringer, and M.~J. Cree.
\newblock {Regularisation of neural networks by enforcing Lipschitz
  continuity}.
\newblock {\em Machine Learning}, 110:393–416, 2021.

\bibitem{Hidano21}
S.~Hidano, T.~Murakami, and Y.~Kawamoto.
\newblock {TransMIA: Membership Inference Attacks Using Transfer Shadow
  Training}.
\newblock In {\em Proc. of IJCNN}, 2021.

\bibitem{Ji18}
Y.~Ji, X.~Zhang, S.~Ji, X.~Luo, and T.~Wang.
\newblock {Model-Reuse Attacks on Deep Learning Systems}.
\newblock In {\em Proc. of CCS}, pages 349--363, 2018.

\bibitem{Jing21}
L.~Jing and Y.~Tian.
\newblock {Self-Supervised Visual Feature Learning with Deep Neural Networks: A
  Survey}.
\newblock {\em IEEE Transactions on Pattern Analysis and Machine Intelligence},
  43(11):4037--4058, 2021.

\bibitem{Kim20}
Y.~Kim, Y.~Kwon, H.~Chang, and M.~C. Paik.
\newblock {Lipschitz Continuous Autoencoders in Application to Anomaly
  Detection}.
\newblock In {\em Proc. of AISTATS}, 2020.

\bibitem{Kurakin17}
A.~Kurakin, I.~J. Goodfellow, and S.~Bengio.
\newblock {Adversarial Examples in the Physical World}.
\newblock In {\em Proc. of ICLR}, 2017.

\bibitem{LeCun98}
Y.~LeCun.
\newblock The mnist database of handwritten digits.
\newblock In {\em http://yann.lecun.com/exdb/mnist/}, 1998.

\bibitem{Liu21}
X.~Liu, F.~Zhang, Z.~Hou, L.~Mian, Z.~Wang, J.~Zhang, and J.~Tang.
\newblock {Self-supervised Learning: Generative or Contrastive}.
\newblock {\em IEEE Transactions on Knowledge and Data Engineering}, page
  DOI:10.1109/TKDE.2021.3090866, 2021.

\bibitem{Liu15}
Z.~Liu, P.~Luo, X.~Wang, and X.~Tang.
\newblock {Deep Learning Face Attributes in the Wild}.
\newblock In {\em Proc. of ICCV}, 2015.

\bibitem{Ng14}
H.-W. Ng and S.~Winkler.
\newblock {A data-driven approach to cleaning large face datasets}.
\newblock In {\em Proc. of ICIP}, pages 343--347, 2014.

\bibitem{Noroozi18}
M.~Noroozi, A.~Vinjimoor, P.~Favaro, and H.~Pirsiavash.
\newblock {Boosting Self-Supervised Learning via Knowledge Transfer}.
\newblock In {\em Proc. of CVPR}, pages 9359--9367, 2018.

\bibitem{Rezaei20}
S.~Rezaei and X.~Liu.
\newblock {A Target-Agnostic Attack on Deep Models: Exploiting Security
  Vulnerabilities of Transfer Learning}.
\newblock In {\em Proc. of ICLR}, 2020.

\bibitem{Salem20}
A.~Salem, A.~Bhattacharya, M.~Backes, M.~Fritz, and Y.~Zhang.
\newblock {Updates-Leak: Data Set Inference and Reconstruction Attacks in
  Online Learning}.
\newblock In {\em Proc. of USENIX Security Symposium}, 2020.

\bibitem{Salem19}
A.~Salem, Y.~Zhang, M.~Humbert, M.~Fritz, and M.~Backes.
\newblock {ML-Leaks: Model and Data Independent Membership Inference Attacks
  and Defenses on Machine Learning Models}.
\newblock In {\em Proc. of NDSS}, 2019.

\bibitem{Shokri17}
R.~Shokri, M.~Stronati, C.~Song, and V.~Shmatikov.
\newblock {Membership Inference Attacks against Machine Learning Models}.
\newblock In {\em Proc. of IEEE Symposium on Security and Privacy}, 2017.

\bibitem{Shorten19}
C.~Shorten and T.~M. Khoshgoftaar.
\newblock {A Survey on Image Data Augmentation for Deep Learning}.
\newblock {\em Journal of Big Data}, 6(60):1--48, 2019.

\bibitem{Simonyan15}
K.~Simonyan and A.~Zisserman.
\newblock {Very deep convolutional networks for large-scale image recognition}.
\newblock In {\em Proc. of ICLR}, 2015.

\bibitem{Sinha21}
A.~Sinha, K.~Ayush, J.~Song, B.~Uzkent, H.~Jin, and S.~Ermon.
\newblock {Negative Data Augmentation}.
\newblock In {\em Proc. of ICLR}, 2021.

\bibitem{Wang18}
B.~Wang, Y.~Yao, B.~Viswanath, and H.~Zheng.
\newblock {With Great Training Comes Great Vulnerability: Practical Attacks
  against Transfer Learning}.
\newblock In {\em Proc. of USENIX Symposium on Security}, pages 1281--1297,
  2018.

\bibitem{Wang20TSC}
S.~Wang, S.~Nepal, C.~Rudolph, M.~Grobler, S.~Chen, and T.~Chen.
\newblock {Backdoor Attacks against Transfer Learning with Pre-trained Deep
  Learning Models}.
\newblock {\em IEEE Transactions on Services Computing}, page DOI:
  10.1109/TSC.2020.3000900, 2020.

\bibitem{Wang19}
W.~Wang, V.~W. Zheng, H.~Yu, and C.~Miao.
\newblock {A Survey of Zero-Shot Learning: Settings, Methods, and
  Applications}.
\newblock {\em ACM Transactions on Intelligent Systems and Technology},
  10(2):13:1--13:37, 2019.

\bibitem{Wang18CVPR}
Y.~Wang, R.~Girshick, M.~Hebert, and B.~Hariharan.
\newblock {Low-Shot Learning from Imaginary Data}.
\newblock In {\em Proc. of CVPR}, pages 7278--7286, 2018.

\bibitem{Wang20}
Y.~Wang, Q.~Yao, J.~T. Kwok, and L.~M. Ni.
\newblock {Generalizing from a Few Examples: A Survey on Few-shot Learning}.
\newblock {\em ACM Computing Surveys}, 53(3):63:1--63:34, 2020.

\bibitem{Weiss16}
K.~Weiss, T.~M. Khoshgoftaar, and D.~Wang.
\newblock {A Survey of Transfer Learning}.
\newblock {\em Journal of Big Data}, 3(1), 2016.

\bibitem{Yang19}
Z.~Yang, J.~Zhang, E.~Chang, and Z.~Liang.
\newblock {Neural Network Inversion in Adversarial Setting via Background
  Knowledge Alignment}.
\newblock In {\em Proc. of CCS}, 2019.

\bibitem{Yao19}
Y.~Yao, H.~Li, H.~Zheng, and B.~Y. Zhao.
\newblock {Latent Backdoor Attacks on Deep Neural Networks}.
\newblock In {\em Proc. of CCS}, pages 2041--2055, 2019.

\bibitem{Ye15}
D.~Ye and M.~Zhang.
\newblock {A self-adaptive strategy for evolution of cooperation in distributed
  networks}.
\newblock {\em IEEE Transactions on Computers}, 64(4):899--911, 2015.

\bibitem{Yin21}
H.~Yin, A.~Mallya, A.~Vahdat, J.~M. Alvarez, J.~Kautz, and P.~Molehanov.
\newblock {See through Gradients: Image Batch Recovery via GradInversion)}.
\newblock In {\em Proc. of CVPR}, pages 16337--16346, 2021.

\bibitem{Yun19}
S.~Yun, D.~Han, S.~J. Oh, S.~Chun, J.~Choe, and Y.~Yoo.
\newblock {CutMix: Regularization Strategy to Train Strong Classifiers with
  Localizable Features}.
\newblock In {\em Proc. of ICCV}, pages 6023--6032, 2019.

\bibitem{Zhang20CVPR}
Y.~Zhang, R.~Jia, H.~Pei, W.~Wang, B.~Li, and D.~Song.
\newblock {The Secret Revealer: Generative Model-Inversion Attacks against Deep
  Neural Networks}.
\newblock In {\em Proc. of CVPR}, pages 253--261, 2020.

\bibitem{Zhang20}
Y.~Zhang, R.~Jia, H.~Pei, W.~Wang, B.~Li, and D.~Song.
\newblock {The Secret Revealer: Generative Model-Inversion Attacks against Deep
  Neural Networks}.
\newblock In {\em Proc. of CVPR}, pages 253--261, 2020.

\bibitem{Zhao21}
X.~Zhao, W.~Zhang, X.~Xiao, and B.~Lim.
\newblock {Exploiting Explanations for Model Inversion Attacks}.
\newblock In {\em Proc. of ICCV}, pages 682--692, 2021.

\bibitem{Zhou16}
B.~Zhou, A.~Khosla, A.~Lapedriza, A.~Oliva, and A.~Torralba.
\newblock {Learning Deep Features for Discriminative Localization}.
\newblock In {\em Proc. of CVPR}, pages 2921--2929, 2016.

\bibitem{Zhuang21}
F.~Zhuang, Z.~Qi, K.~Duan, D.~Xi, and Y.~Zhu.
\newblock {A Comprehensive Survey on Transfer Learning}.
\newblock {\em Proceedings of the IEEE}, 109(1):43--76, 2021.

\bibitem{Zou20}
Y.~Zou, Z.~Zhang, M.~Backes, and Y.~Zhang.
\newblock {Privacy Analysis of Deep Learning in the Wild: Membership Inference
  Attacks against Transfer Learning}.
\newblock In {\em https://arxiv.org/abs/2009.04872}, 2020.

\end{thebibliography}

\end{document}